\documentclass[11pt,letterpaper]{article}
\pdfoutput=1
\usepackage{jcappub}
\usepackage{color}
\usepackage{graphicx}

\usepackage{verbatim}
\usepackage{amsmath}
\usepackage{amssymb}
\usepackage{subfig}
\usepackage{url}
\usepackage{bbold}

\usepackage{multirow}
\usepackage{threeparttable}
\usepackage{paralist}

\newcommand{\SU}{\text{SU}}
\newcommand{\SO}{\text{SO}}
\newcommand{\Sp}{\text{Sp}}

\newcommand{\U}{\text{U}}

\DeclareRobustCommand{\Sec}[1]{Sec.~\ref{#1}}

\DeclareRobustCommand{\Tab}[1]{Table~\ref{#1}}

\DeclareRobustCommand{\Fig}[1]{Fig.~\ref{#1}}

\DeclareRobustCommand{\Eq}[1]{Eq.~(\ref{#1})}

\newcommand{\be}{\begin{equation}}
\newcommand{\ee}{\end{equation}}
\newcommand{\bea}{\begin{eqnarray}}
\newcommand{\eea}{\end{eqnarray}}

\newcommand{\mb}[1]{\boldsymbol{#1}}

\usepackage{xspace}

\def\Tr{\mathop{\rm Tr}}

\begin{document}

\title{Dark Nuclei I: Cosmology and Indirect Detection}
 
\author{William Detmold,}
\author{Matthew McCullough,}
\author{and Andrew Pochinsky}

\affiliation{Center for Theoretical Physics, Massachusetts Institute of Technology, Cambridge, MA 02139, USA}

\emailAdd{wdetmold@mit.edu}
\emailAdd{mccull@mit.edu}
\emailAdd{avp@mit.edu}

\abstract{In a companion paper (to be presented), lattice field theory methods are used to show that in two-color, two-flavor QCD there are stable nuclear states in the spectrum.  As a commonly studied theory of composite dark matter, this motivates the consideration of possible nuclear physics in this and other composite dark sectors.  In this work, early Universe cosmology and indirect detection signatures are explored for both symmetric and asymmetric dark matter, highlighting the unique features that arise from considerations of dark nuclei and associated dark nuclear processes.  The present day dark matter abundance may be composed of dark nucleons and/or dark nuclei, where the latter are generated through {\it dark nucleosynthesis}.  For symmetric dark matter, indirect detection signatures are possible from annihilation, dark nucleosynthesis, and dark nuclear capture and we present a novel explanation of the galactic center gamma ray excess based on the latter.  For asymmetric dark matter, dark nucleosynthesis may alter the capture of dark matter in stars, allowing for captured particles to be processed into nuclei and ejected from the star through dark nucleosynthesis in the core.  Notably, dark nucleosynthesis realizes a novel mechanism for indirect detection signals of asymmetric dark matter from regions such as the galactic center, without having to rely on a symmetric dark matter component.}

\preprint{MIT-CTP {4554}}

\maketitle

\section{Introduction}
\label{sec:introduction}
It remains a pressing challenge in particle physics to understand the particle nature of dark matter (DM).  The relentless experimental exploration of the possible interactions between DM and Standard Model (SM) fields has revealed a great deal of crucial information about potential interactions.  However, as yet no unambiguous signals of DM have emerged, and many popularly considered DM candidates have come under increasing pressure from null experimental results.  This situation motivates the continued, and ever-diversifying experimental and theoretical efforts to probe the DM frontier.  In particular, it is pertinent to map out the theoretical landscape of DM paradigms, as candidates with exotic properties may motivate the consideration of non-standard experimental signatures of DM.  In recent years, there has been a surge of interest in models of DM with distinctive interactions and/or multiple states.  Along these lines, the properties of the SM fields have in some cases guided the exploration of possibilities for the dark sector via analogy.  Popular examples include dark sectors, or sub-sectors, with dark atomic behavior \cite{Alves:2010dd,Behbahani:2010xa,Kaplan:2011yj,Kumar:2011iy,Khlopov:2011tn,Cline:2012is,CyrRacine:2012fz,Fan:2013yva,Fan:2013tia,McCullough:2013jma,Cline:2013pca,Belotsky:2014haa} or strongly-coupled dark sectors leading to composite DM candidates \cite{Nussinov:1985xr,Barr:1990ca,Khlopov:2005ew,Gudnason:2006ug,Gudnason:2006yj,Khlopov:2008ty,Ryttov:2008xe,Foadi:2008qv,Alves:2009nf,Mardon:2009gw,Kribs:2009fy,Frandsen:2009mi,Lisanti:2009am,Khlopov:2010pq,Belyaev:2010kp,Lewis:2011zb,Buckley:2012ky,Hietanen:2012qd,Hietanen:2012sz,Appelquist:2013ms,Hietanen:2013fya,Cline:2013zca,Appelquist:2014dja}, which are the focus of this work.\footnote{See also \cite{Braaten:2013tza,Laha:2013gva} for treatment of annihilation and scattering dynamics in composite dark sectors where resonant effects are important.}

Composite DM, which arises due to confining gauge dynamics in the dark sector, has been considered for some time.  In all studies thus far, the DM candidate has been assumed to be hadron of the dark sector, such as a dark meson or a dark baryon.  However, if the analogy with the SM is taken seriously there is also the possibility of stable composites of the hadrons themselves: {\it dark nuclei}.  The nuclei of the SM provide a clear proof-of-principle that such states may exist.  Determining the spectrum of nuclei in any strongly coupled gauge theory is a difficult task, only now becoming possible through advances in the application of lattice field theory methods \cite{Beane:2012vq,Detmold:2012eu,Yamazaki:2012hi}. This explains why, thus far, only the hadronic spectrum of postulated strongly coupled dark sectors has been studied seriously.  In a first step towards quantitatively exploring the possibility of a dark nuclear spectrum, we will present lattice calculations in a companion paper that demonstrate that in two-color, two-flavor QCD, stable nuclear states are possible with the lowest lying states being a bound states of $\pi$ and $\rho$ mesons and their baryonic partners.  Thus any discussion of DM candidates in this theory now necessitate some consideration of the nuclear states. Going further, this suggests the possibility of analogues of nuclei should be considered in {\it any} strongly interacting composite model.   Our work substantially extends, and is complementary to, earlier pioneering lattice studies of DM candidates in such strongly coupled sectors \cite{Lewis:2011zb,Hietanen:2012qd,Hietanen:2012sz,Appelquist:2013ms,Hietanen:2013fya,Appelquist:2014dja}.

As will be demonstrated, the phenomenology of dark sectors exhibiting composite DM candidates broadens significantly when the possibility of dark nuclei is introduced.  In this work, we construct a model based on the broad qualitative findings of the lattice study and undertake an exploration of the cosmology and possible indirect detection signatures of dark nuclei.

The genesis of dark nuclei is achieved through a {\it dark nucleosynthesis} processes.\footnote{Some aspects of dark nucleosynthesis have been discussed in Ref.~\cite{Krnjaic:2014xza} that appeared as we were concluding our study.}  A prototypical example in the SM is the first step of nucleosynthesis, $n+p \to d + \gamma$, where $d$ is a deuteron.  For symmetric dark sectors, {\it dark nuclear capture} is also possible, and an analogous SM example would be $\overline{p}+d \to n + \gamma$.  Generally speaking, the broad topology of both processes is that of so-called {\it semi-annihilation} \cite{D'Eramo:2010ep,D'Eramo:2011ec,D'Eramo:2012rr,Belanger:2012vp}, which has also arisen in other models \cite{Arina:2009uq,Hambye:2009fg,Hambye:2008bq}.  We will find that the distinguishing features of dark nucleosynthesis arise from the small binding energies involved in these reactions ({\it i.e.}, in the SM, $M_d \simeq M_n + M_p$).  In the case of asymmetric DM, the conservation of dark baryon-number also leads to novel possibilities.  For symmetric and asymmetric DM, the early Universe cosmology may be altered quite radically by dark nucleosynthesis, and in extreme cases it is possible that the interactions are strong enough such that all the available dark nucleons may  be processed into dark nuclei through a late period of dark nucleosynthesis, much as the available SM neutrons are processed into nuclei in Big Bang Nucleosynthesis.

The phenomenology of indirect detection may also be modified significantly.  This is most notable for the case of asymmetric DM.  In standard asymmetric DM scenarios, indirect detection signals are not possible unless some symmetric DM component is present. This effectively makes the indirect detection signature a feature of symmetric, rather than asymmetric, DM.  However, dark nucleosynthesis preserves dark baryon number and is thus possible for a purely asymmetric dark sector.  If the additional neutral states produced in dark nucleosynthesis are observable, this leads to a novel mechanism for the indirect detection of asymmetric DM. Again, this may be seen through the analogous SM nucleosynthesis process, $n+p \to d + \gamma$.  In the case of symmetric DM, the usual DM annihilation processes are possible, however the new channels of dark nucleosynthesis and dark nuclear capture may give rise to additional signals.  Furthermore, the energy scale associated with dark nucleosynthesis is hierarchically smaller than that of annihilation, and this may lead to complementary signals from the same DM candidates that would have the same spatial morphology, but at very different energy scales.

The phenomenology of DM capture in stars and other astrophysical bodies may also be significantly altered by dark nucleosynthesis.  DM may become captured within stars, with a rate determined by the magnitude of the DM-nucleon scattering cross section.  If the DM is asymmetric, then dark nucleosynthesis may lead to indirect detection signatures, in contrast to standard asymmetric DM candidates.  Furthermore, even for relatively small binding energy fractions, dark nucleosynthesis may result in the dark nucleus being ejected from the Sun, or other bodies.  This hinders the buildup of asymmetric DM within stars, leading to significantly different phenomenology from the signatures of standard asymmetric DM.

In \Sec{sec:lattice}, we will briefly review the lattice field theory calculations which provide evidence for the presence of stable nuclear states in two-color, two-flavor QCD, leaving the full technical details to the companion paper.  In \Sec{sec:model}, we present a simplified model of the dark sector based on dark $\pi$, $\rho$, fields as well as dark nuclei $D$ (for simplicity, we restrict our discussion to the lightest dark nucleus) and a dark Higgs, $h_D$.  This simple effective theory serves to mock-up the qualitative (though not necessarily quantitative) behavior of the relevant states and interactions, allowing for an exploration of the particle phenomenology.  In \Sec{sec:cosmo}, we solve the relevant Boltzmann equations to determine the relic abundance of the dark nucleons and dark nuclei for various interaction strengths for both symmetric and asymmetric DM scenarios.  In \Sec{sec:indirect}, we explore the indirect detection signatures of the model.  In \Sec{sec:galcentex} we discuss a novel explanation of the galactic center gamma ray excess based on dark nuclear capture.  In \Sec{sec:indirectnucleo} we present a novel paradigm for asymmetric DM indirect detection through dark-baryon number conserving nucleosynthesis reactions and we briefly sketch potential modifications of the phenomenology of DM capture in stars which arise due to the introduction of dark nucleosynthesis, leaving detailed study to a dedicated analysis. We conclude in \Sec{sec:conclusions}.

\section{Lattice investigation of model spectrum}
\label{sec:lattice}

In this work, we focus on a putative model for dark matter involving a strongly interacting SU($N_c=2$) gauge theory with $N_f=2$ degenerate fermions in the fundamental representation. In a companion paper, we undertake a detailed, lattice field-theoretic  exploration of the spectroscopy of hadronic states that appear in this model.  Importantly, we show that light stable nuclei (systems with baryon number $B\geq2$) appear even in  this simple model and we extract the spectrum of the lightest few nuclei for representative values of the fermion masses. In this section, we summarize the main results that are obtained from these calculations. 

As will be discussed below, this model  has a large set of global symmetries that constrain the dynamics in the limit of
vanishing quark masses. It is expected that the theory produces five degenerate (pseudo-)Goldstone boson states: three mesons analogous to the usual QCD pions, and a baryon and anti-baryon which are (pseudo-)Goldstone bosons  carrying baryon number. Ref.~\cite{Buckley:2012ky} considered the interesting possibility that dark baryon number is  conserved 
and that dark matter is composed of the Goldstone baryon with a mass parametrically small compared to typical strong interactions in the theory that are set by the scale $\Lambda_{N_c=2}$. 
In our numerical investigations, we focus on a regime of the model in which explicit chiral symmetry breaking through 
quark masses is dominant over the effects of dynamical chiral symmetry breaking. This regime is characterised by having $0.5 < M_\pi/M_\rho < 1$, where $M_\pi$ and $M_\rho$ are the masses of particle in the lightest multiplets containing pseudoscalar and vector mesons (and their baryon partners), respectively. 

After a careful analysis of the relevant correlation functions of this theory at multiple lattice spacings and multiple volumes, we are able to extract the continuum limit, infinite volume spectrum of light nuclei for a range of relevant quark masses. While there is some variation with the quark masses that are used, the overall picture that emerges from these calculations is as follows. 
\begin{itemize}
\item Spin $J=1$ axial-vector nuclei with baryon number $B=2$ and $3$ are clearly bound, with energies below the threshold for breakup into individual baryons. The $J=1$, $B=4$ system is likely bound, but our results are not precise enough to be definitive in this case. Higher baryon number states with $J=1$ are clearly above the relevant breakup thresholds and do not form bound states.
\item The binding energies of these systems are quite deep. Measured in units of the dark pion decay constant, $f_\pi$, we find dimensionless binding energies per baryon $\Delta E_B/B f_\pi\sim0.1$ (for SM nuclei, the same quantity ranges between 0.01 and 0.06). Phrased in terms of the individual baryon masses, the bindings are at the few percent level.
For different values of the quark masses, the precise values of these ratios  will change and both smaller and larger values of the binding seem feasible.
\item Spin $J=0$ scalar multi-baryon systems are probably not bound states (although the systematic uncertainties are somewhat large in this case). Baryons with higher spin and in different flavour representations have not been studied.
\item By performing calculations with a range of quark masses, the Feynman-Hellmann theorem can be used to extract the $\sigma$-terms for the various hadrons that govern the couplings of the states of the theory to scalar currents. These couplings 
are found to be of a natural size, with  $f_{q}^{(H)} \equiv \frac{\langle H | m_q \overline{q}q|H\rangle}{M_H}\sim 0.15$--$0.3$.
\end{itemize}
Full details of the calculations and results will be presented in the companion paper.
In principle, lattice field theory methods  can also be used to investigate elastic scattering in the dark sector and provide determinations of couplings of the dark states to an analogue electroweak sector and/or to other parts of the dark sector. However, such calculations are beyond our current scope, and we will instead rely on dimensional analysis and these qualitative results to provide estimates in our discussion of the rich phenomenology of this theory.

\section{An Explicit Model of Dark Nuclei}
\label{sec:model}

Building upon these lattice investigation, a demonstrative model of dark nucleosynthesis is now presented.

\subsection{Dark Mesons}

\begin{table}[t]
\centering
\begin{tabular}{ c | c | c | c | c | c}
Field & Spin & $\SU(2)_L$ &  $\SU(2)_R$ \\
\hline \hline
$u_L$ & $1/2$ & $\mb{\Box}$ & $\mb{1}$ \\
$u_R$ & $1/2$ & $\mb{1}$ & $\mb{\Box}$ \\
$d_L$ & $1/2$ & $\mb{\Box}$ & $\mb{1}$ \\
$d_R$ & $1/2$ & $\mb{1}$ & $\mb{\Box}$ \\
$H_D$ & $0$ & $\mb{1}$ & $\mb{1}$ \\
\end{tabular}
\caption{Field content and gauge interactions of the model in the UV.}
\label{tab:content}
\end{table}

The field content of the model is shown in \Tab{tab:content} and the Lagrangian is
\be
\mathcal{L} = \mathcal{L}_{\rm strong} - \frac{\lambda}{4} \left(v_D- H_D^2 \right)^2 - \left( \kappa H_D (u_R^\dagger u_L + d_L^\dagger d_R) + h.c. \right) ~.
\ee
The strong dynamics of the SU($N_c=2$) sector is described implicitly within $\mathcal{L}_{\rm strong}$ and characterized by a scale $\Lambda_{\rm QC_2D}$.
 $H_D$ is a `dark' Higgs boson as this model could be UV completed in such a way that $h_D$ is the Higgs boson remaining after spontaneous symmetry breaking of a dark $\U(1)$ gauge symmetry.   We assume that $v_D$ and the scalar quartic- and Yukawa-interactions are sufficiently small that the resulting dark Higgs boson mass and the quark masses are below the strong coupling scale $m_{h_D} \lesssim \Lambda$.  Approaching the strong coupling scale from above, the relevant interactions are
\be
\mathcal{L} = \mathcal{L}_{\rm strong} - V (h_D) - \left( m_q(1 + h_D/v_D) (u_R^\dagger u_L + d_L^\dagger d_R) + h.c. \right) ~,
\ee
which includes the $\SU(2)_D$ gauge interactions.  In the absence of the Yukawa terms and quark masses, there is an $\SU(2)_L \times \SU(2)_R$ global symmetry which is enlarged to $\SU(2)_L \times \SU(2)_R \rightarrow \SU(4)$  because the $\SU(2)_D$ representations are pseudo-real, enabling the right-handed quarks to fall into multiplets alongside the left-handed quarks.  

We also include a small mixing term between the visible-sector Higgs boson and the dark Higgs boson through the Higgs portal operator $|h_D|^2 |H|^2$.  The dark Higgs boson is a SM gauge singlet, hence below the scale of $\U(1)_D$ breaking, this coupling mimics the usual mixing between a SM singlet scalar and the SM Higgs boson.  This is introduced to enable the dark Higgs to decay via standard Higgs boson decay channels such as $h_D \to \overline{b} b$.  There are already strong constraints on the allowed mixing angle, and we thus assume this mixing is small, below the $\sim \text{few} \%$ level \cite{Bertolini:2012gu,Belanger:2013kya,Giardino:2013bma,Ellis:2013lra}.

Below the strong-coupling scale, a quark condensate forms and breaks the global symmetry  $\SU(4) \rightarrow \Sp(4)$ \cite{Peskin:1980gc,Preskill:1980mz,Kosower:1984aw}.  There are five pseudo-Goldstone bosons corresponding to the broken generators of $\SU(4)$.  They obtain mass due to the quark mass terms which break this symmetry explicitly.  Three of these pseudo-Goldstone bosons are familiar from QCD and can be thought of as the pions ($\pi^0,\pi^+,\pi^-$) made up of the u- and d-quarks and anti-quarks.  The other two pseudo-Goldstone bosons may be thought of as $u d$ and $\overline{u} \overline{d}$ composites carrying baryon-number.  We denote these pseudo-Goldstone bosons as $\pi^B$ and $\pi^{\overline{B}}$.  Thus there are in total five pseudo-Goldstone degrees of freedom denoted $\pi^0,\pi^+,\pi^-,\pi^B,\pi^{\overline{B}}$.

As with the analogous QCD case, the Goldstone manifold for $\SU(4)/\Sp(4)$ may be parameterized as
\be
\Sigma = U \Sigma_c U^T
\ee
where 
\be
U = \exp \left[ \frac{i}{f} \left( \begin{array}{cccc} \pi^0 & \sqrt{2} \pi^+ & 0 & \sqrt{2} \pi^B \\ \sqrt{2} \pi^- & -\pi^0 &  -\sqrt{2} \pi^B & 0 \\  0 & - \sqrt{2} \pi^{\overline{B}} & \pi^0 & \sqrt{2} \pi^- \\   \sqrt{2} \pi^{\overline{B}} & 0 &  \sqrt{2} \pi^- & -\pi^0 \end{array} \right) \right], \qquad   \text{and  } \Sigma_c  = \left( \begin{array}{cccc} 0 & 0 & -1 & 0 \\ 0 & 0 & 0 & -1 \\ +1 & 0 &0 & 0 \\ 0 & +1 & 0 & 0 \end{array} \right).
\ee
Under chiral rotations, $\Sigma \to L \Sigma R^\dagger$ (where $L$ and $R$ are rotations in the underlying SU(2)$_{L,R}$), or equivalently, $\Sigma \to G \Sigma G^\dagger$ where $G$ is an SU(4) rotation.
The quark mass matrix can be written as $M_q = m_q (1 + h_D/v_D) \Sigma_c$ and may be thought of as transforming under $\SU(4)$ in the same way as the pion field $\Sigma$.  The pion masses and Higgs-pion couplings may be determined from the $\SU(4)$-invariant chiral Lagrangian
\be
\mathcal{L}_{eff} = \frac{f^2}{2} \Tr \partial_\mu \Sigma \partial^\mu \Sigma^\dagger - G_\pi m_q (1 + h_D/v_D) \Tr (\Sigma_c \Sigma) ~~,
\ee
where $G_\pi$ is an unknown dimensionful constant.  As all pions are equally massive, they  couple to the Higgs in the same way.

There are also five vector mesons which are odd under the analogue of $G$-parity.  Since we choose $m_q$ comparable to the strong scale, they have similar masses to the pions.  We continue the analogy with QCD and denote these vector bosons $\rho^0_\mu,\rho^+_\mu,\rho^-_\mu,\rho^B_\mu,\rho^{\overline{B}}_\mu$, with the latter two carrying baryon number +1 and -1, respectively.  The vector bosons and their interactions with the pseudo-Goldstone bosons are constrained by chiral symmetry. This can be implemented in  a number of ways including through the `heavy-field' formalism \cite{Coleman:1969sm,Callan:1969sn,Jenkins:1995vb}, since the mass of these particles remains nonzero even for vanishing quark masses.  Using this approach, we introduce
$\xi=\sqrt{\Sigma}$ (transforming as $\xi \to \sqrt{R\Sigma L^\dagger}$) and parameterize the vector boson fields as a $4\times4$ matrix of fields, $O_\mu$, in analogy with the pion fields.

The leading interactions are parameterized with the Lagrangian
\bea
{\cal L}_{v}&=& -i {\rm tr}\left[O^\dagger_\mu {\cal V}\cdot {\cal D} O^\mu\right]  
+i g_{V} {\rm tr}\left[\{O^\dagger_\mu,O_\nu\} {\cal A}_\lambda\right] v_\sigma \epsilon^{\mu\nu\rho\sigma}   \nonumber \\
&&+ M_{V,0}^2 {\rm tr}\left[O^\dagger_\mu O^\mu\right] 
 +\lambda_1{\rm tr} \left[\{O^\dagger_\mu,O^\mu\} {\cal M}\right]  
 +\lambda_2{\rm tr} \left[O^\dagger_\mu O^\mu\right] {\rm tr} \left[{\cal M}\right]  
\eea
where ${\cal M}= \frac{1}{2}(\xi m_q \xi + \xi^\dagger m_q \xi^\dagger)$ and ${\cal D}^\mu O^\nu=\partial^\mu O^\mu +[{\cal V}^\mu,O^\nu]$ with  ${\cal V}^\mu=\frac{1}{2}(\xi \partial^\mu \xi^\dagger +\xi^\dagger\partial^\mu\xi)$ and ${\cal A}^\mu=\frac{i}{2}(\xi \partial^\mu \xi^\dagger -\xi^\dagger\partial^\mu\xi)$. $g_V$, $\lambda_1$ and $\lambda_2$ are unknown dimensionless couplings and $M_{V,0}\sim \Lambda_{\rm QC_2D}$ is the vector boson mass in the limit of vanishing quark masses.

We will only need the lowest-order couplings of the dark Higgs to the composite bosons, and express them as 
\be \label{eq:hmes}
\mathcal{L}_{\text{Int}} =  A_\pi h_D \left((\pi^0)^2/2 + \pi^+ \pi^- + \pi^B \pi^{\overline{B}} \right)+ A_\rho h_D \left( (\rho^0)^2/2 + \rho^+ \rho^- + \rho^B \rho^{\overline{B}} \right) ~,
\ee
where the sum over Lorentz indices for the vector mesons is implied.  This completes the interactions necessary for the annihilation processes $\pi \pi \to h_D h_D$ and $\rho \rho \to h_D h_D$ relevant for the cosmological abundance and indirect detection signals of these states.  We will take these couplings to be free parameters in what follows, however for a specific choice of quark masses they could be calculated from the $\sigma$-terms discussed in \Sec{sec:lattice} where it is found that the couplings take perturbative values of $\mathcal{O}(0.1)$.  With DM masses near the weak scale this suggests that annihilation and nucleosynthesis cross sections would typically take weak-scale values.

\subsection{Dark Nuclei}
As demonstrated through the lattice calculation, in this simple model a $\pi$ boson and a 
$\rho$ boson may combine to form stable two-body bound states: the dark nucleus, $D$.  
In analogy to the visible sector, we will refer to the $\pi$ and $\rho$ bosons as dark nucleons, and to the $D$ as the dark deuteron. These dark nuclei have mass $M_D = M_\pi + M_\rho - B_D$ where $B_D$ is the binding energy of the dark nucleus and may take a range of values.  In what follows we will assume the isospin symmetric case where any of the five dark $\pi$ bosons may combine with any of the five dark $\rho$ bosons, leading to a total of $25$ dark nuclei which may carry dark baryon number $Q_B=0,\pm1,\pm2$.   Although the lattice calculations give specific values for the binding energies, we do not wish to restrict ourselves to particular values of masses, binding energies, and coupling constants. We thus allow these to be free parameters throughout,  taking the lattice values as a rough guide.  We will only consider dark nuclei composed of two dark nucleons in order to simplify the treatment of the cosmology and indirect detection phenomenology. The lattice calculations suggest that three- and perhaps four-body states may also be stable, which would enrich the phenomenology even further. Other possibile examples of strongly interacting dynamics may produce higher-body bound states as well.

Assuming $m_{h_D} < B_D$, dark nucleosynthesis proceeds in this model via the process $\pi + \rho \to D+ h_D$, in analogy with the first step of nucleosynthesis in the Standard Model,  $n + p \to d+ \gamma$.  As discussed in \Sec{sec:introduction}, the reaction $\pi + \rho \to D + h_D$ is a semi-annihilation reaction as the number of dark matter states changes by one $(\text{stable}+\text{stable}) \to (\text{stable}+\text{unstable})$, followed by $(\text{unstable}) \to (\text{SM})$.  In this work we call this particular realization of semi-annihilation {\it dark nucleosynthesis} to reflect that dark nuclei are forming from dark nucleons.

In order to estimate the cosmological relic abundance of the dark nuclei, or the indirect detection signals from dark nucleosynthesis, it is necessary to determine the dark nucleosynthesis cross section $\sigma (\pi \rho \to D h_D)$.  A full nuclear effective field theory estimation would treat the dark nuclear scattering amplitude as an infinite sum of dark nucleon loops and determine the corresponding propagator for the dark nucleus from this sum.  Such a treatment is well beyond the scope of this work and instead we opt for a simplified effective field theory estimation which takes the rudimentary assumption of treating the dark nucleus as a fundamental state at energies near or below $m_D$.  Effective operators for $\pi$, $\rho$, and $D$ interactions are then determined from the symmetry structure and dimensional analysis.

In terms of the remaining $\Sp(4)$ global flavour symmetry, the $\pi$ and $\rho$ fields both live in the coset space $\SU(4)/\Sp(4)$.    Rather than constraining the interactions using an $\Sp(4)$ basis for the $D$ fields, we instead  utilize the local isomorphism $\Sp(4) \cong \SO(5)$.   The $\pi$ and $\rho$ bosons transform as fundamentals under the the global $\SO(5)$ symmetry.  Thus the $D$ fields, which are composites of these two fundamentals, must decompose as the tensor product $\mb{5} \times \mb{5} = \mb{1} + \mb{10} + \mb{14}$.  These $\SO(5)$ representations are at most $2$-index, simplifying the calculation of vertices relative to the alternative $\Sp(4)$ representations.  The bosons in $\SO(5)$ are real degrees of freedom and do not fall naturally into the classification of pions and baryons discussed above.  However, the two bases for these fields may be simply found from the following unitary rotation $\mb{\pi} = U \cdot \mb{\pi}_R$, where the subscript $R$ denotes a real $\SO(5)$ representation.  Specifically, this relationship is 
\be \label{eq:rotate}
 \left( \begin{array}{c} \pi^+  \\   \pi^- \\   \pi^0  \\   \pi^B \\   \pi^{\overline{B}} \end{array} \right) = \frac{1}{\sqrt{2}} \left( \begin{array}{ccccc} +1 & +i & 0 & 0& 0 \\ +1 & -i &  0 & 0& 0 \\ 0 & 0 &  +\sqrt{2} & 0 & 0 \\  0 & 0 & 0 & +1 & +i  \\  0 & 0& 0&+1 & -i  \end{array} \right) \cdot \left( \begin{array}{c} \pi_1  \\   \pi_2 \\   \pi_3  \\   \pi_4 \\   \pi_5 \end{array} \right) ~~,
\ee
and similarly for the $\rho$ mesons.  The $25$ real degrees of freedom in $D$ furnish an $\SO(5)$ singlet, an antisymmetric representation, and a symmetric representation.  Using the rotation of \Eq{eq:rotate}, we may relate this basis of real fields to a more intuitive basis of $5$ real and $10$ complex vector fields which have varying baryon number.  This representation is
\be \label{eq:Dmu}
\mb{D}^\mu =  \left( \begin{array}{ccccc} 
S^\mu_+ & D^\mu_{2,0} & D^\mu_{1,0} & D^\mu_{1,-1} & D^\mu_{1,1} \\ 
\overline{D}^\mu_{2,0} & S^\mu_- &  D^\mu_{-1,0} & D^\mu_{-1,-1} & D^\mu_{-1,1} \\ 
\overline{D}^\mu_{1,0} & \overline{D}^\mu_{-1,0} &  S^\mu_0 &  D^\mu_{0,-1} & D^\mu_{0,1} \\  
\overline{D}^\mu_{1,-1} & \overline{D}^\mu_{-1,-1} & \overline{D}^\mu_{0,-1} & S^\mu_B & D^\mu_{0,2}  \\  
\overline{D}^\mu_{1,1}& \overline{D}^\mu_{-1,1} & \overline{D}^\mu_{0,1}+ & \overline{D}^\mu_{0,2} & S^\mu_{\overline{B}}  \end{array} \right)  ~~,
\ee
where all diagonal elements are real and the subscript denotes the states that the diagonal elements couple to in the notation of the pion fields.  The off-diagonal elements are complex vectors for which the first subscript denotes the global $\U(1)_D$ charge and the second subscript the dark $\U(1)_B$ baryon number in the same units as the pions.  In this notation the various real $\SO(5)$ representations may be written as
\begin{eqnarray}
\mb{D}^\mu_{\mb{1}} ~& = & \Tr (\mb{D}^\mu) ~,\\
\mb{D}^\mu_{\mb{10}} & = & \frac{i}{2} \left( \mb{D}^\mu-\mb{D}^{\mu T} \right)  ~, \\
\mb{D}^\mu_{\mb{14}} & = & \frac{1}{2} \left( \mb{D}^\mu+\mb{D}^{\mu T} \right) - \frac{1}{5} \Tr (\mb{D}^\mu) \mathbb{1}_5 ~.
\end{eqnarray}

\begin{figure}[]
  \centering
  \includegraphics[height=0.22\textwidth]{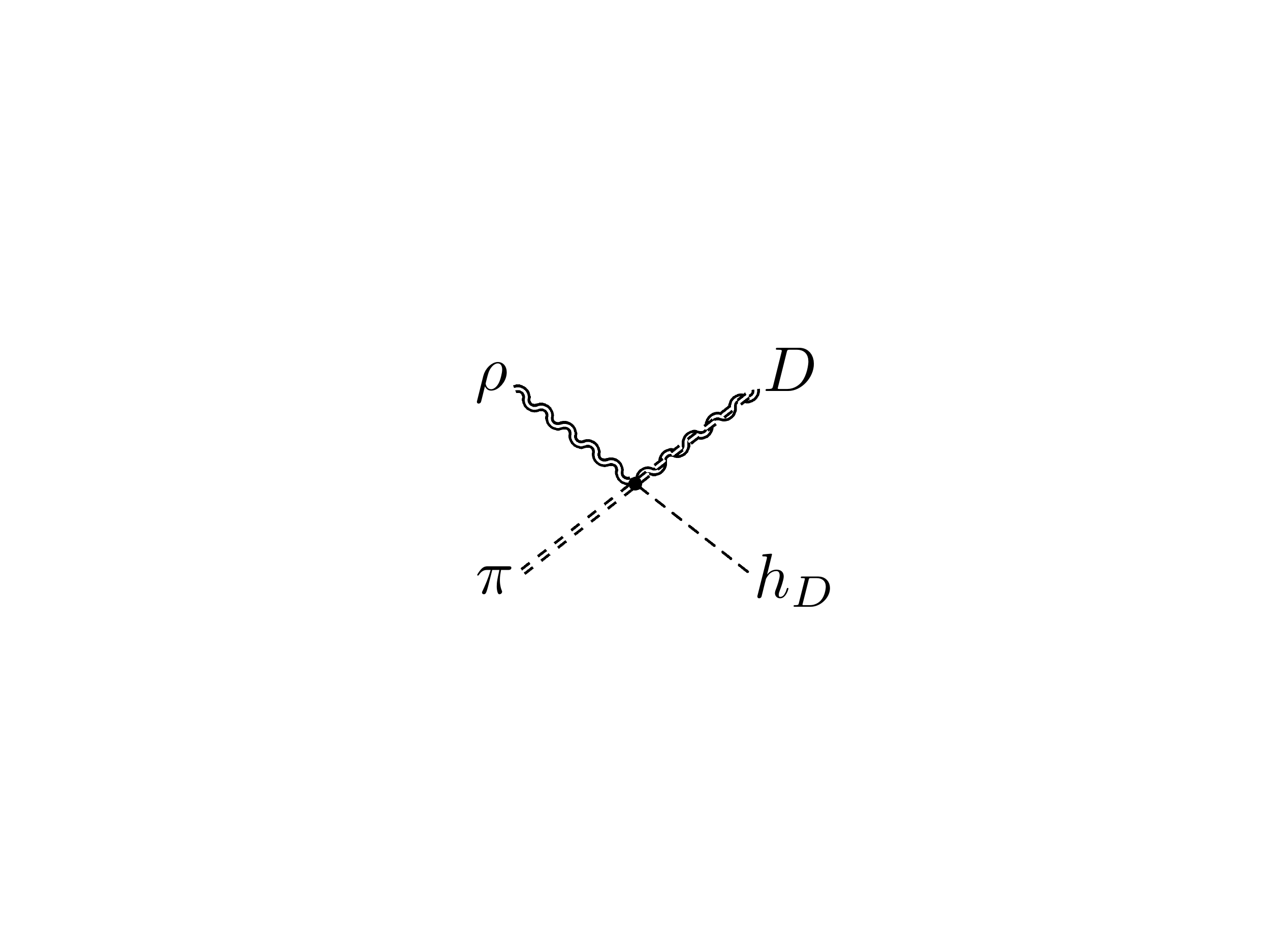}
  \caption{A dark nucleosynthesis event.  This is realized in the model of \Sec{sec:model} and is analogous to the SM process $n+p \to D + \gamma$.  Such dark nucleosynthesis processes are important in early Universe cosmology as they may alter relic abundances.  In the present day they may also be relevant as they may give rise to observable indirect detection signatures from the galactic center and from stars.}
  \label{fig:darknucdiag}
\end{figure}

The lattice calculation considered the nuclei in the symmetric representation, $\mb{D}_{\mb{14}}$, finding bound states for a range of quark masses, but did not investigate the singlet or antisymmetric representations. To simplify the calculations relevant for phenomenology, we will assume that all nuclei representations are stable and equally massive.  This is purely for the sake of simplifying the phenomenology, however if it turned out that the antisymmetric represent were unstable this would only result in minor modifications.  There is some contribution to the mass of the dark nuclei from the  masses of the constituent hadrons, and some from their interactions.  For the regime in which it makes sense to call $D$ a `nucleus',  the binding energy should be small, $B_D \ll M_\pi, M_\rho$, and the first contribution from the constituent masses should to be dominant. Since the nuclei are ultimately built from  quarks, there is a coupling to the Higgs field which we may write (under the assumption of equal masses) as
\be \label{eq:hD}
\mathcal{L}_{\text{Int}} =  \frac{1}{2} A_D h_D \Tr \left(\mb{D}^\dagger \mb{D} \right)\,,
\ee
where again $A_D$ is taken as a free parameter of $\mathcal{O} (0.1 \times \Lambda_{QC_2 D})$.  Also, consistent with the remaining symmetries in the real-field basis the $\mb{1}$, $\mb{10}$, and $\mb{14}$ of $\SO(5)$ may couple to the mesons as 
\be
\label{eq:Lrpd}
\mathcal{L}_{\rho\pi D} \sim \mb{\pi}^\dagger (\bar\lambda_{\mb{1}} \mb{D}^\mu_{\mb{1}}+\bar\lambda_{\mb{10}} \mb{D}^\mu_{\mb{10}}+\bar\lambda_{\mb{14}} \mb{D}^\mu_{\mb{14}}) \mb{\rho}_{\mu}\,.
\ee
The remaining symmetry does not constrain these interactions any further, however to simplify the calculation of annihilation and semi-annihilation cross sections we make the further additional assumption that $\bar\lambda_{\mb{1}}=\bar\lambda_{\mb{10}}=\bar\lambda_{\mb{14}}=\bar\lambda$, thus the coupling written in terms of the real degrees of freedom may be simply expressed as $\mathcal{L}_{\pi\rho D} = \bar\lambda \mb{\pi}_R^\dagger \cdot \mb{D}^\mu_R \cdot \mb{\rho}_{\mu R}$ where $\mb{D}_R$ is a $5\times5$ matrix of real fields.  This trilinear coupling, combined with the dark Higgs couplings, leads to dark nucleosynthesis, $\pi + \rho \to D + h_D$, by dressing one of the external propagators in three-body scattering with a dark Higgs vertex.  If all parameters were known, then these additional couplings and diagrams should be included in a full treatment of semi-annihilation.  However, as the energy carried away by $h_D$ in the semi-annihilation process is $E_{h_D} \sim \mathcal{O} (B_D) \ll M_\pi, M_\rho, m_D$, we may integrate out these interactions to generate an effective quartic vertex
\be \label{eq:vertex}
\mathcal{L}_{\rm Eff} = \lambda h_D \mb{\pi}_R^\dagger \cdot \mb{D}_R^\mu \cdot \mb{\rho}_{\mu R}~,
\ee
where $\lambda$ is taken as a free parameter assumed to be $\lambda \sim \mathcal{O}(0.1)$.  This interaction is depicted in \Fig{fig:darknucdiag}.  There would also be an effective quartic vertex of the same form simply from the effective theory and this additional contribution is absorbed into the parameter $\lambda$.  Thus, \Eq{eq:hmes} and \Eq{eq:hD} contain all of the information relevant for annihilation, and \Eq{eq:vertex} determines  dark nucleosynthesis.

\section{Cosmology of Dark Nucleosynthesis}
\label{sec:cosmo}
The cosmology and possible experimental signatures of dark nuclei, and in particular of dark nucleosynthesis, are rich subjects. Throughout we aim to stress the differences between scenarios with dark nuclei and standard dark matter models, finding that dark nuclei may possess a very distinctive phenomenology.  We will appeal to the specific model of \Sec{sec:model} in order to illustrate the signatures.  We do this to demonstrate that explicit realizations of these signatures exist, and also for the pedagogical purposes of providing a familiar example. However, we emphasize that the signatures are common to the broad class of possibilities for dark nuclei and are not restricted to this model.  As such, the various cross sections are taken as free parameters and, motivated by the values of the $\sigma$-terms determined from the lattice calculation, they are assumed to be $\sigma \sim \mathcal{O}(0.1^2/8 \pi M_\pi^2)$.  We begin by considering the early Universe cosmology and relic abundance of a sector capable of dark nucleosynthesis.

\subsection{Symmetric Dark Matter}\label{sec:freeze}
Thermal freeze-out of the coupled system involves the $\pi$ and $\rho$ nucleons and $D$ nuclei of \Sec{sec:model}.  For a symmetric DM scenario, it is useful to return to the real basis of fields.  This is because all $5$ $\pi$ meson degrees of freedom are equally massive and similarly for the $5$ $\rho$ mesons and the $25$ nuclei.  We will also use the rotated form of the nucleus matrix such that all of these fields are contained within a $5\times5$ matrix of real fields where each field interacts with a particular $\pi$ and $\rho$ combination in the same way. The assumed symmetry reduces the coupled system of Boltzmann equations down from $35$ individual equations to $3$ as the number density of any $\pi^a$ must be equal to the number density of any other $\pi^b$ and so on for the other fields.  We thus write $n_{\pi^a} = n_{\pi}/5$, $n_{\rho^a} = n_{\rho}/5$, $n_{D^a} = n_{D}/25$.  Also, the total number of $\pi$ degrees of freedom is $5$, the total number of $\rho$ degrees of freedom is $5\times 3 = 15$ due to the spin states of the massive vectors, and for the nuclei, there are  $25 \times 3 = 75$ degrees of freedom.  For simplicity, we will also assume that $M_\pi = M_\rho$.

\begin{figure}[t]
  \centering
 \includegraphics[height=0.45\textwidth]{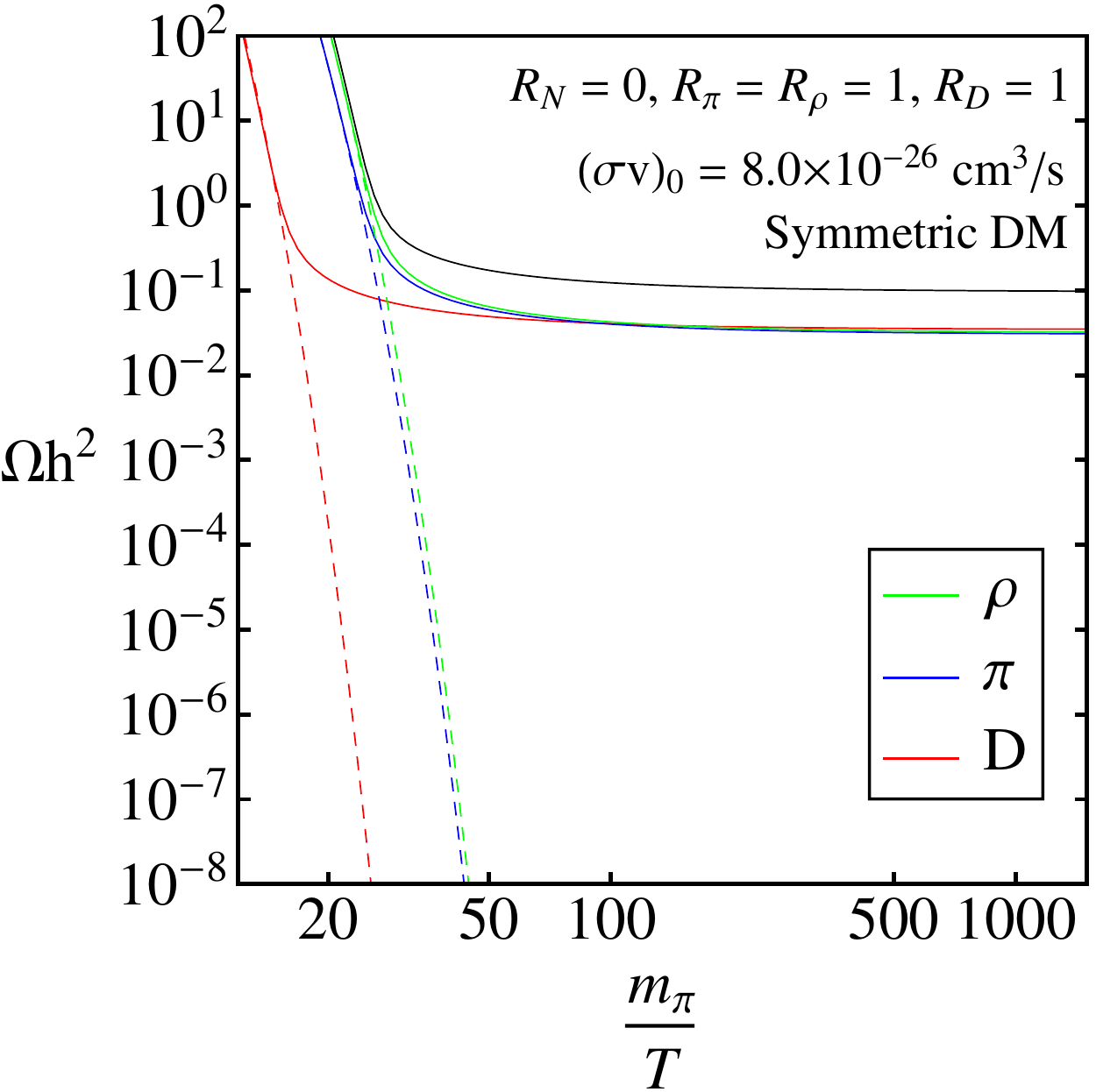} \qquad  \includegraphics[height=0.45\textwidth]{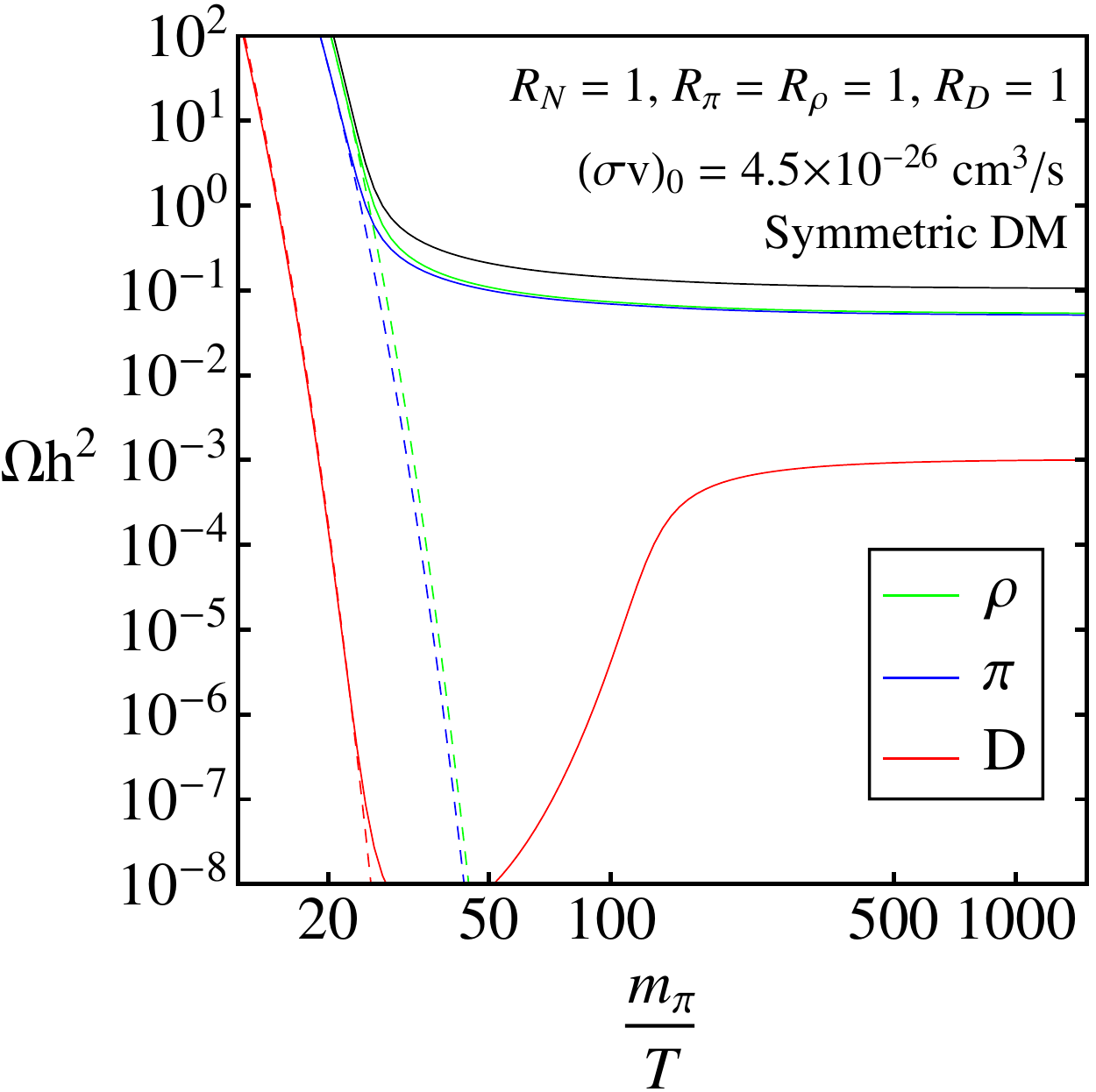} \\
 \includegraphics[height=0.45\textwidth]{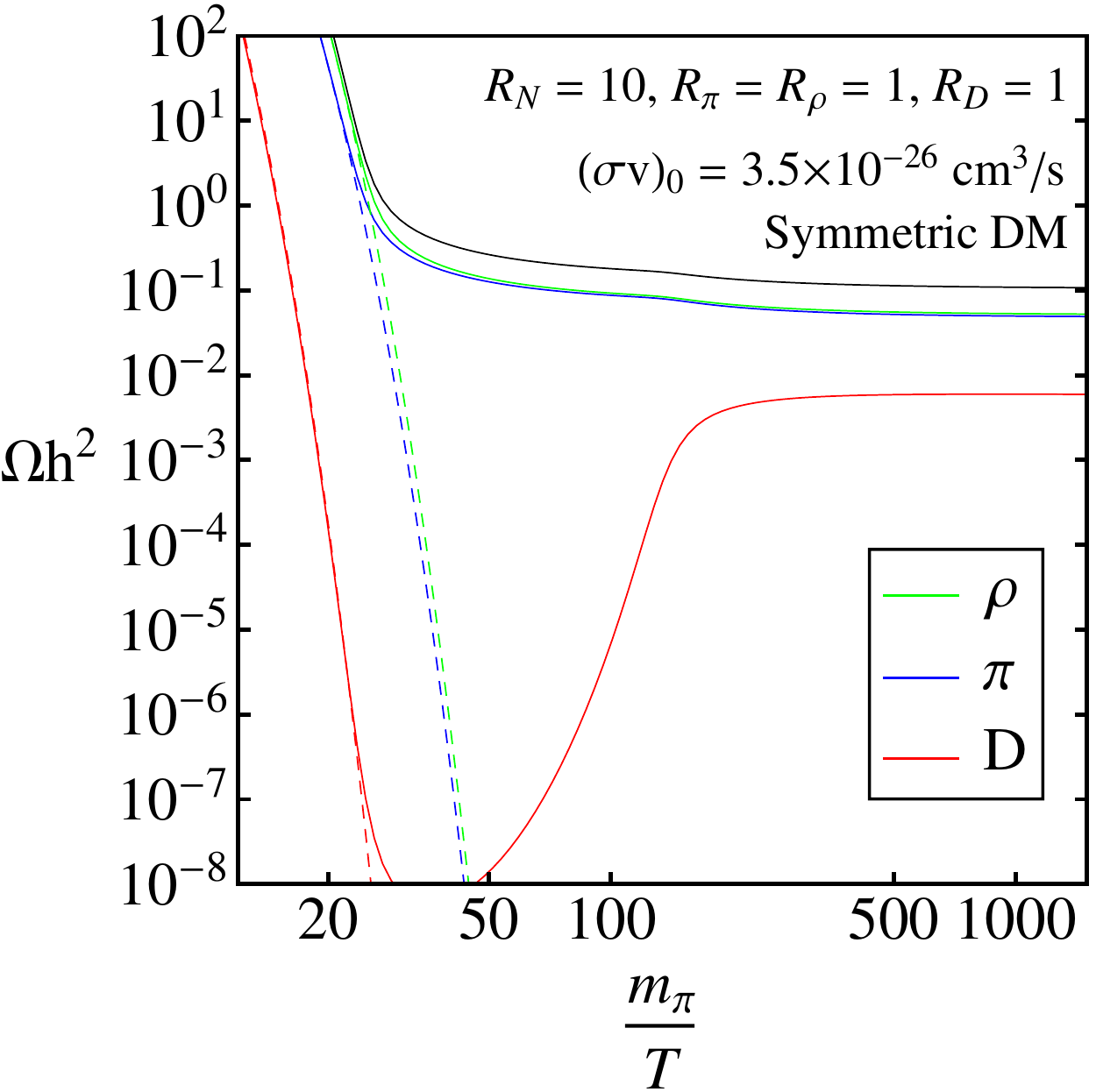}\qquad  \includegraphics[height=0.45\textwidth]{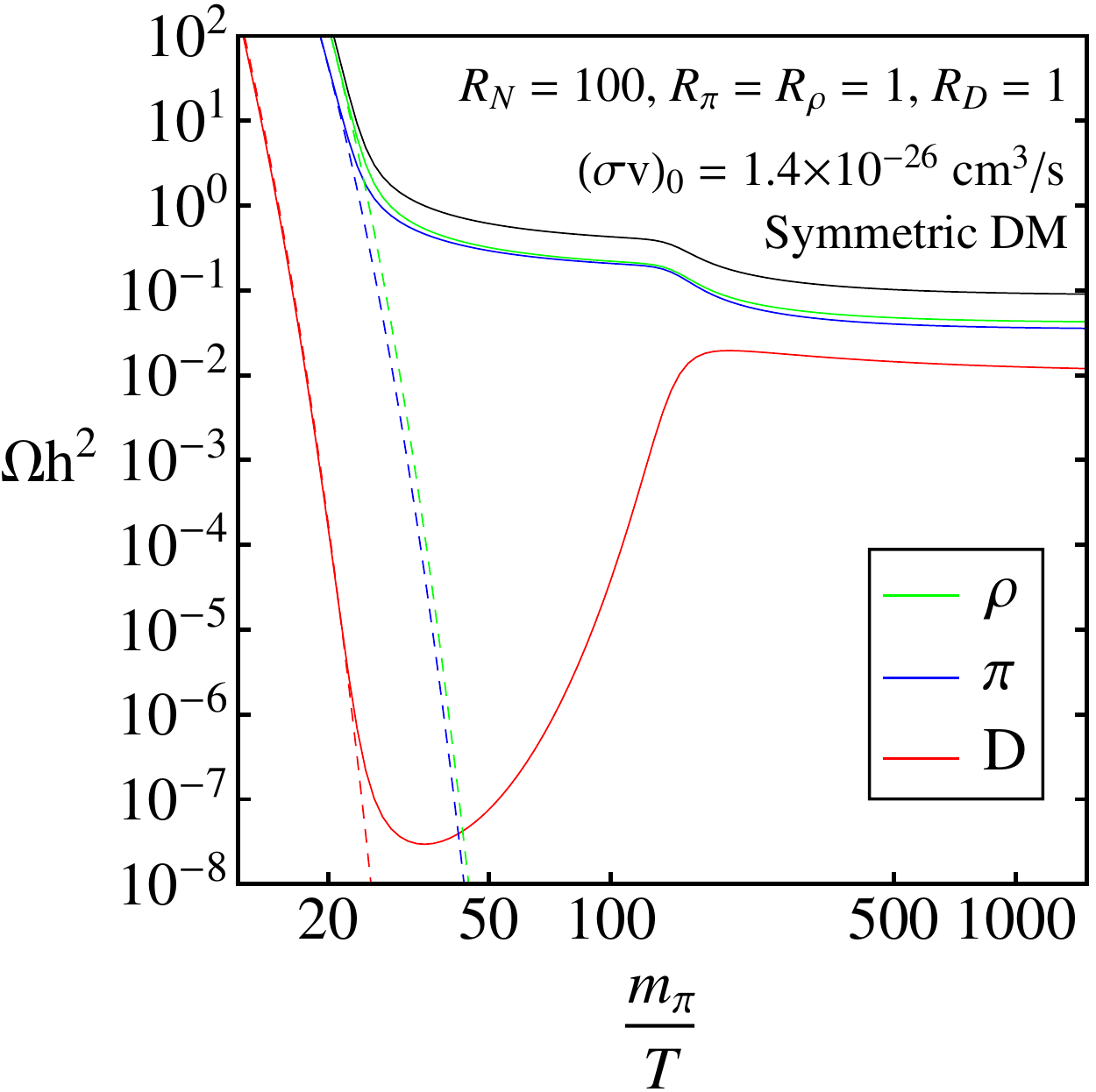}
  \caption{Relic density of nucleons and nuclei in the presence of annihilations and dark nucleosynthesis.  Nucleon masses are $M_\pi = M_\rho = 100$ GeV, the dark Higgs at $10$ GeV, and the binding energy fraction $\delta=0.1$ ($B=10$ GeV), thus dark nucleosynthesis occurs precisely at threshold.  The full solutions are shown as solid lines and the equilibrium values as dashed lines.  The total DM abundance is shown in solid black.  Even a small dark nucleosynthesis cross section may have a dramatic effect on the relic density, most notably as the nuclei may remain in thermal equilibrium through interactions with nucleons down to the freeze-out temperature of the lighter nucleons.  Interestingly once all of the nuclei and nucleons fall out of thermal equilibrium the nucleus fraction may be repopulated at lower temperatures due to the continued nucleosynthesis reactions.}
  \label{fig:thermal}
\end{figure}

If we let $(\sigma v)_0$ be a free parameter describing the typical scale for scattering cross sections in the dark sector which is of order the weak scale, we may write the thermally-, and spin-averaged individual dark nuclear capture cross section as $\langle \sigma v (\pi^a D^b \to \rho^c h_D) \rangle =\langle \sigma v (\rho^a D^b \to \pi^c h_D) \rangle = R_N (\sigma v)_0$ where the subscript denotes that this is a nuclear process.\footnote{Note that this particular capture process only occurs for specific combinations of nucleons and nuclei, for example $\pi_{\overline{B}} + D_{0,2} \to \rho_B + h_D$, while other channels are excluded.} If we write the nuclear binding energy as $B_D = \delta\, M_\pi$ and the dark Higgs boson mass as $M_{h_D}=\kappa\, M_\pi$, the dark nucleosynthesis process $\pi^a+ \rho^b \to D^c +h_D$ is only possible at zero relative velocity if $\kappa < \delta$.  Even in this case, dark nucleosynthesis must occur close to the kinematic threshold.  It was shown some time ago that in determining the cosmological evolution of DM abundances, any near-threshold processes have distinctive features when compared to more typical processes, such as annihilation to light states \cite{Griest:1990kh}.  In order to simplify the presentation of results in this section we choose $\kappa = \delta$ in many instances, such that dark nucleosynthesis may only occur exactly on threshold.  We have not found an analytic solution for the thermally averaged cross section in the most general case, and hence choose to provide an approximate expression.  For the case where $\delta>\kappa$ and nucleosynthesis is possible at zero relative velocity, we calculate the standard velocity-independent cross section.  To this, we include 
the thermally averaged cross section when nucleosynthesis is possible exactly on threshold ($\delta = \kappa$) which we calculate following Ref.~\cite{Griest:1990kh}.  The resulting expression is approximate, however it is appropriate for the case we will usually consider with $\delta = \kappa$, and has the correct limits in the more general case.  Thus we find that the thermally-averaged nucleosynthesis cross section is
\begin{eqnarray}
\langle \sigma v (\pi^a \rho^b \to D^c h_D) \rangle & \approx &  \frac{9}{4} \left( \sqrt{\delta^2-\kappa^2} + \frac{3}{\sqrt{\pi x}} \left( 1-\frac{4}{3 x} \right) \right) R_N (\sigma v)_0 \nonumber\\
& = & f(x) R_N (\sigma v)_0~,
\label{eq:threshold}
\end{eqnarray}
where $x=M_\pi/T$, in agreement with the results of \cite{Griest:1990kh}.  As expected, this cross section vanishes in the zero temperature limit at threshold ($\delta=\kappa$) and if nucleosynthesis is kinematically allowed ($\delta>\kappa$) the correct limit is reached for s-wave scattering in the zero temperature limit.  The various spin-averaged annihilation cross sections may be parameterized relative to $(\sigma v)_0$ as
\begin{eqnarray}
\langle \sigma v (\pi^a \pi^a \to h_D h_D) \rangle/5  = R_\pi (\sigma v)_0~,\nonumber \\
\label{eq:sigma0}
\langle \sigma v (\rho^a \rho^a \to h_D h_D) \rangle/15  = R_\rho (\sigma v)_0~,\\
\langle \sigma v (D^a D^a \to h_D h_D) \rangle/75  = R_D (\sigma v)_0~, \nonumber
\end{eqnarray}
where $R_\pi$, $R_\rho$ and $R_D$ are simple rescaling factors introduced to allow different annihilation cross sections for the various fields.  The co-moving number densities are written as $Y_{\pi,\rho,D} = n_{\pi,\rho,D}/s$, where $n_a$ is the temperature-dependent number density of a particle species and $s$ is the temperature-dependent entropy density. The equilibrium co-moving number densities are defined as $Y_f^{eq}$ and we use the parameterization
\be
\lambda = \frac{5 x (\sigma v)_0}{H(M_\pi)} \bigg|_{x=1} .
\ee
With all of these definitions in place the set of coupled Boltzmann equations for all particle species may be rearranged following standard methods \cite{Kolb:1990vq} and are written
\begin{eqnarray}
\frac{dY_\pi}{dx} & = & - \lambda \bigg[R_\pi \left(Y_\pi^2 - {Y_\pi^{eq}}^2 \right) +\frac{1}{5}  R_N (Y_\pi Y_D - \frac{Y_\rho}{Y_\rho^{eq}} Y_\pi^{eq} Y_D^{eq}) \\
&&-\frac{1}{5}  R_N(Y_\rho Y_D - \frac{Y_\pi}{Y_\pi^{eq}} Y_\rho^{eq} Y_D^{eq})+  R_N f(x) (Y_\pi Y_\rho - \frac{Y_D}{Y_D^{eq}} Y_\pi^{eq} Y_\rho^{eq})  \bigg]~, \nonumber \\
\frac{dY_\rho}{dx} & = & - \lambda \bigg[R_\rho \left(Y_\rho^2 - {Y_\rho^{eq}}^2 \right) + \frac{1}{5}  R_N(Y_\rho Y_D - \frac{Y_\pi}{Y_\pi^{eq}} Y_\rho^{eq} Y_D^{eq})\nonumber \\
&& -\frac{1}{5}  R_N (Y_\pi Y_D - \frac{Y_\rho}{Y_\rho^{eq}} Y_\pi^{eq} Y_D^{eq})+ R_N f(x) (Y_\rho Y_\pi - \frac{Y_D}{Y_D^{eq}} Y_\rho^{eq} Y_\pi^{eq})  \bigg]~, \nonumber \\
\frac{dY_D}{dx} & = & - \lambda \bigg[R_D \left(Y_D^2 - {Y_D^{eq}}^2 \right) -R_N f(x) (Y_\pi Y_\rho - \frac{Y_D}{Y_D^{eq}} Y_\pi^{eq} Y_\rho^{eq})  \nonumber  \\
&&  + \frac{1}{5} R_N \left((Y_\pi + Y_\rho ) Y_D - \left(\frac{Y_\rho}{Y_\rho^{eq}} Y_\pi^{eq}+ \frac{Y_\pi}{Y_\pi^{eq}} Y_\rho^{eq} \right) Y_D^{eq} \right) \bigg]~,\nonumber
\label{eq:Boltzmann}
\end{eqnarray}
where the various multiplicities of the species have been taken into account.  Further, in any given nucleosynthesis reaction the symmetry structure requires that only one nucleus is produced for any particular combination of $\pi$ and $\rho$.  This can be seen clearly in the $\SO(5)$ basis.  These coupled Boltzmann equations may then be solved to determine the total relic abundance of dark matter, and also the relative abundances of the dark nucleons, $\rho,\pi$, and the dark nuclei $D$.  The energy density in any particle relative to the critical density may be determined from the particle mass and the current entropy density.

\Fig{fig:thermal} shows some typical solutions to the Boltzmann equations.  It is clear that dark nucleosynthesis may have a pronounced effect on the final relic density, with the greatest effect coming from the additional destruction of nuclei through the dark nuclear capture processes $\pi^a +D^b \to \rho^c+ h_D$.  It is clarifying to break the evolution of the dark nuclei into a number of smaller steps:
\begin{itemize}
\item $\mb{T > 2 M_\pi / 20}$: The number density of dark nucleons and nuclei tracks the equilibrium density due to efficient annihilations.

\item $\mb{M_\pi / 20 < T < 2 M_\pi / 20}$: The dark nuclei are kept at equilibrium density below the temperature of dark nuclei annihilation freeze out due to efficient dark nuclear capture interactions with the dark nucleons which are themselves still efficiently annihilating.  Freeze out of the dark nuclei is paused until the lighter dark nucleons freeze out, hence the greatly suppressed number density of dark nuclei.  This can be seen from \Fig{fig:thermal} where in cases with dark nucleosynthesis, the freeze out of the dark nuclei is paused until dark nucleon freeze out.

\item $\mb{B_D / 20 < T < M_\pi / 20}$: In this regime, all annihilations have effectively frozen out, and the only remaining interactions are dark nucleosynthesis interactions.  The possible reaction types are nucleosynthesis, $\pi + \rho \to D + h_D$, and nuclear capture, $D + (\pi,\rho) \to h_D + (\rho,\pi)$.\footnote{There may also be capture processes such as $D + \rho \to \rho + h_D$, however these would be p-wave suppressed and thus subdominant to the s-wave processes that we consider.}  The cross section for the former is suppressed due to the reduced phase space, however the interaction rate for the latter is suppressed to a greater degree due to the extremely small number density of dark nuclei.  Hence during this era the dark nuclei effectively `freeze in' \cite{Hall:2009bx} as their number density increases exponentially while the total energy density in DM slowly bleeds off through dark nucleosynthesis.

\item $\mb{T < B_D / 20}$: In this era all reactions, including dark nucleosynthesis, have effectively frozen out and the number density of all species is now fixed
\end{itemize}

This completes our discussion of cosmological evolution of the symmetric scenario for dark nuclei.

\subsection{Asymmetric Dark Matter}
We now consider more directly the analogy with standard nucleosynthesis and consider an asymmetric DM (ADM) scenario.  Recent years have seen a resurgence in the study of ADM \cite{Nussinov:1985xr,Gelmini:1986zz,Chivukula:1989qb,Barr:1990ca,Kaplan:1991ah,Thomas:1995ze,Hooper:2004dc,Kitano:2004sv,Agashe:2004bm,Cosme:2005sb,Farrar:2005zd,Suematsu:2005kp,Tytgat:2006wy,Banks:2006xr,Kitano:2008tk,Kaplan:2009ag}, and this has led to the realization of a large number of models which may generate a DM asymmetry through a variety of mechanisms.\footnote{See \cite{Petraki:2013wwa,Zurek:2013wia} for recent reviews.}  Thus there are many plausible scenarios in which an asymmetry may be generated in the dark sector.  In this work, we will focus on heavy asymmetric DM \cite{Buckley:2010ui,MarchRussell:2011fi} which is a complementary scenario to the usual $M\sim5$ GeV asymmetric DM, however the lighter $M\sim5$ GeV possibility for asymmetric dark nucleons and/or nuclei is equally possible.

Motivated by the analogy with nucleosynthesis, we consider a scenario where the asymmetry in the dark sector is in dark baryon-number, thus $n_{\pi_B} \gg n_{\pi_{\overline{B}}}$.  Also, for the sake of simplicity we will assume that the only relevant fields are the dark Higgs $h_D$, the dark baryon-number carrying mesons $\pi^{B}, \pi^{\overline{B}},\rho^{B}, \rho^{\overline{B}}$ and the dark baryon-number charge $2$ fields $D^{B}$, and $D^{\overline{B}}$ (note the change in notation for convenience).  It may be possible to realize this in a full scenario as an appropriate splitting between quark masses may explicitly break the global $\SU(4)$ symmetry sufficiently that $M_{\pi^\pm} \gg M_{\pi^0}, M_{\pi^{B,\overline{B}}}$.  In turn, this makes all nuclei containing $M_{\pi^\pm}$ heavy as well.  As the $\pi^0$ field is neutral under the remaining global symmetries we may introduce new decay channels for this field, hence the dominant DM phenomenology may be determined by considering only the dark baryon-number $1$ nucleons and dark baryon-number $2$ nuclei.  However, we have chosen to make this assumption primarily to simplify the treatment of the phenomenology.

In order for the relic abundance to be dominated by an asymmetry, the annihilation cross-section for all states must exceed the thermal relic annihilation cross section which, given that the dark sector is strongly coupled, seems plausible.  In this case, the relic abundance of the symmetric component is suppressed by a factor $\sim \exp(\sigma_\text{Ann}/\sigma_{\text{Th}})$ where the latter is the standard thermal DM cross section \cite{Graesser:2011wi}.  This is a result of continued annihilations with the asymmetric component.  With the symmetric DM component mostly annihilated away, the dominant component of DM is comprised of the baryon-number carrying states shown in \Tab{tab:remnants}.

\begin{table}[h]
\centering
\begin{tabular}{ c | c | c | c | c | c | c}
State & $\pi^B$ & $\rho^B$ & $D^{B}$ \\
\hline \hline
Dark Baryon Number & $+1$ & $+1$ & $+2$ \\
\end{tabular}
\caption{Relic DM states carrying dark baryon-number in the asymmetric scenario.}
\label{tab:remnants}
\end{table}

If we consider the production of an asymmetry in dark baryon number in the early Universe, then at later times this asymmetry may be understood by considering the chemical potential for dark baryon number $\mu_D$.  If the dark nucleosynthesis interactions $\pi^B + \rho^B \to D^B + h_D$ are efficient, then we obtain the relationship between chemical potentials $\mu_\pi + \mu_\rho = \mu_{D}$.  Similarly, we will assume that at high temperatures around the strong coupling scale we would have $\mu_\pi = \mu_\rho$, however it is not possible to determine the full details of chemical equilibrium in practice at these scales without evolving through the strong coupling scale.

\begin{figure}[t]
  \centering
 \includegraphics[height=0.45\textwidth]{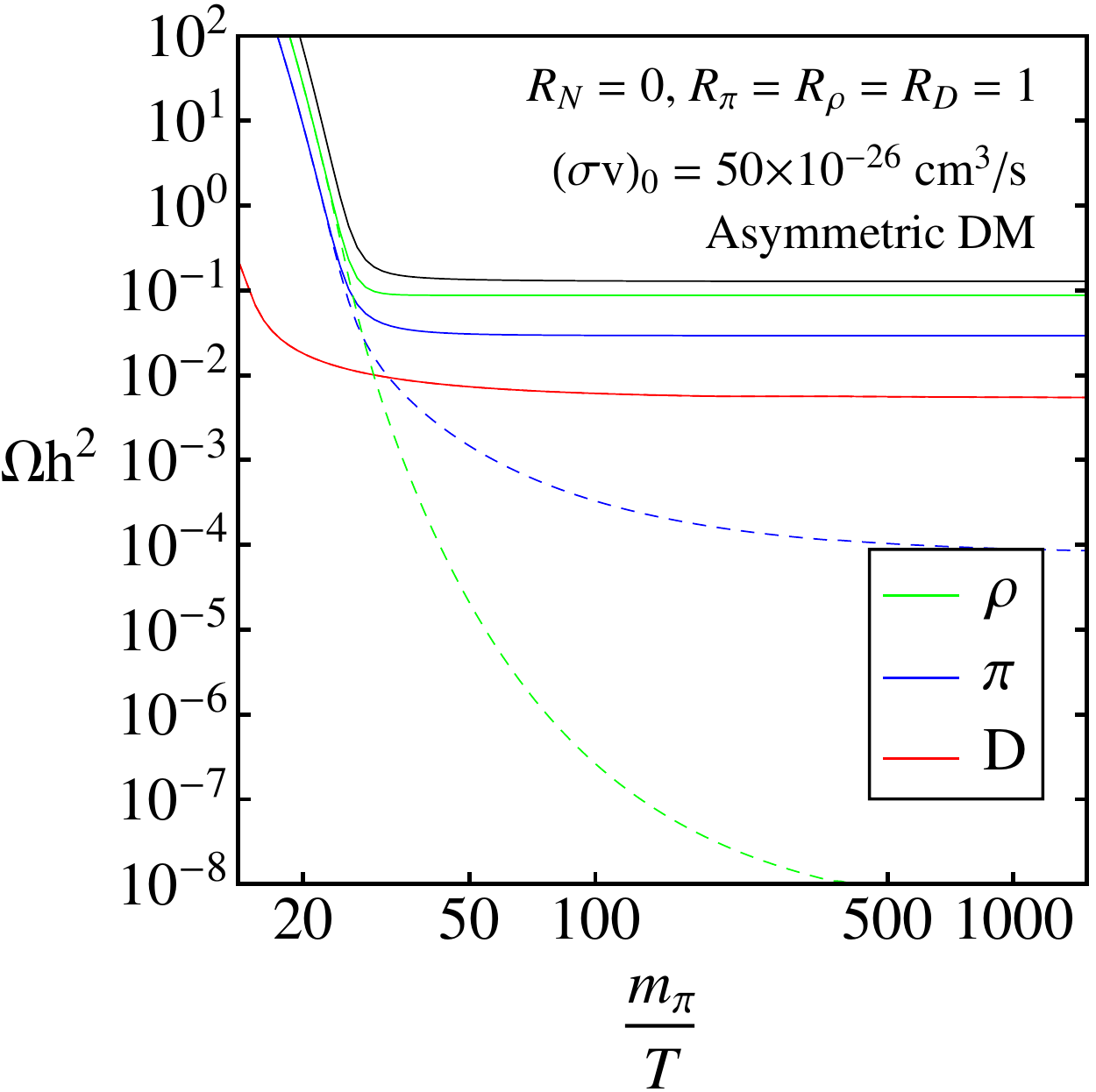} \qquad  \includegraphics[height=0.45\textwidth]{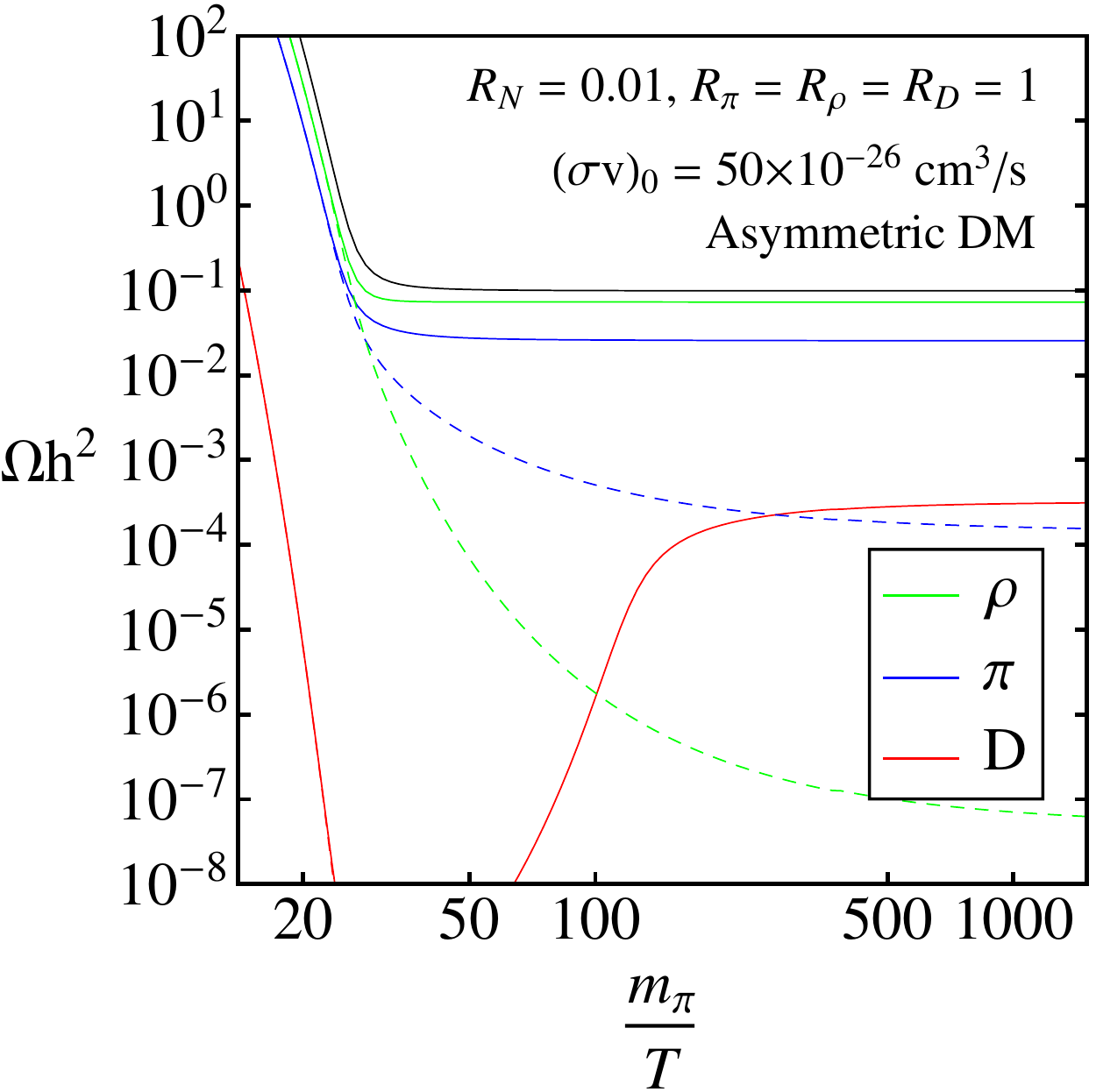} \\
 \includegraphics[height=0.45\textwidth]{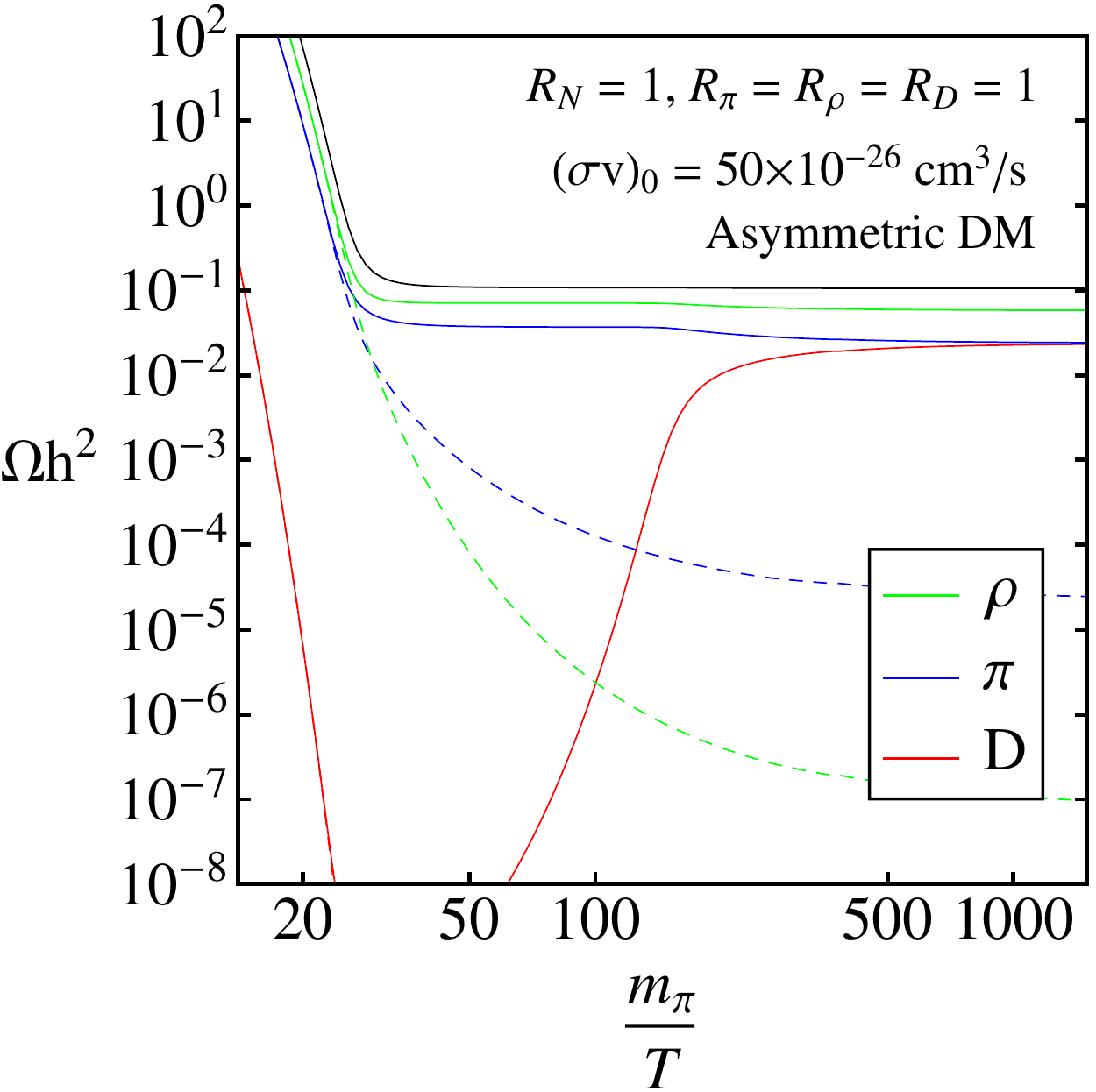}\qquad  \includegraphics[height=0.45\textwidth]{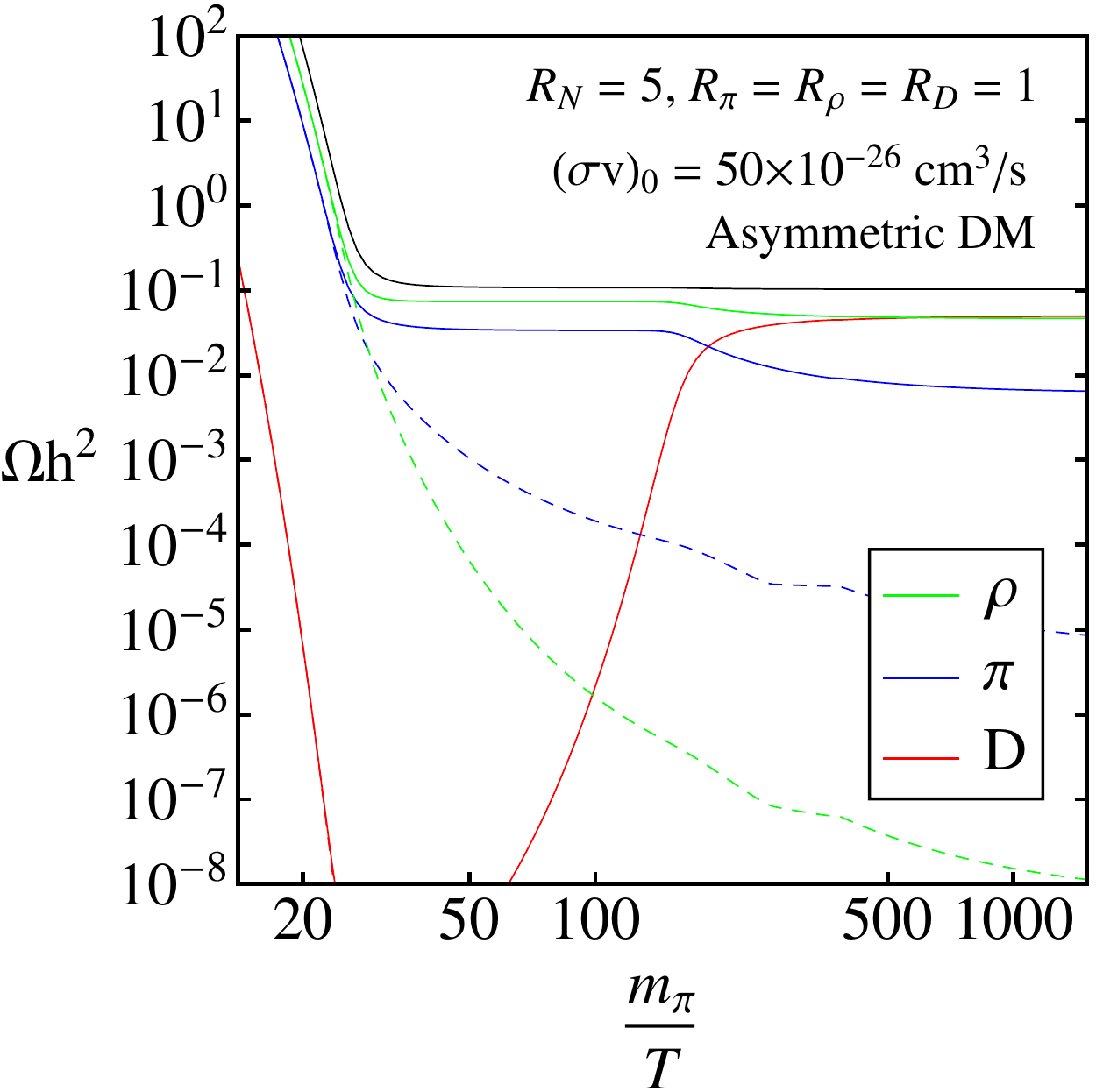}
  \caption{Relic density of dark nucleons and nuclei in the presence of annihilations and dark nucleosynthesis for the case of asymmetric DM.  Nucleon masses are  $M_\pi = M_\rho = 100$ GeV, the dark Higgs at $10$ GeV, and the binding energy fraction $\delta=0.1$, thus dark nucleosynthesis occurs precisely at threshold.  The dark baryon densities are shown as full lines and the anitbaryon densities as dashed lines.  The total DM abundance is shown in solid black.  Once again, dark nucleosynthesis may have a pronounced effect on the relic density of the various species.  Many of the features, including the timeline of the various freeze-out epochs, are similar to the symmetric DM case.  However, due to the preservation of the asymmetry larger dark nucleosynthesis cross sections may be tolerated while maintaining the observed DM abundance, and in this case the majority of available dark nucleons may be processed into dark nuclei.}
  \label{fig:thermalasymm}
\end{figure}

Before considering the Boltzmann equations, it is illuminating to consider general features of dark nucleosynthesis in the asymmetric case.  If we specify the number densities of the various DM species relative to the number density of photons as $\eta_a = n_a/n_\gamma$, we may relate the total asymmetric dark number density to the cosmological abundance of DM by taking the ratio of the known baryon asymmetry and baryon abundance
\be
\eta_{N_D} = \eta_\pi + \eta_\rho + 2 \eta_D \approx 2.68\times 10^{-8} \times \left( \frac{\Omega_{DM} h^2}{\Omega_B h^2} \frac{M_H}{M_\pi} \right),
\ee
where $M_H$ is the mass of hydrogen.  From this, denoting  the fractional asymmetry in a given species as $X_a = Q_{B_a} n_a/ (n_\pi + n_\rho + 2 n_D)$, we have the fractional asymmetry carried in dark nuclei
\be
X_{D} = \frac{1}{3} X_\pi^2 \eta_{N_D} \left( 2-\frac{B_D}{M_\pi} \right)^{3/2} \left( \frac{2 \pi}{M_\pi T} \right)^{3/2} \exp^{B_D/T}~.
\label{eq:DNscheme}
\ee
For temperatures well above the binding energy, $T\gtrsim B_D$, the exponential is small and $X_D \ll X_\pi$.  However, if chemical equilibrium is maintained to temperatures $T\ll B_D$ such that the exponential overcomes the small value of the asymmetry in either $\pi$ or $\rho$, which is $\eta_{N_D} \sim \mathcal{O}(10^{-8})$, then the majority of the asymmetric component will actually be carried in the dark nuclei.  In fact, this is already familiar from nucleosynthesis in the SM where the strong interactions maintain chemical equilibrium to temperatures well below the binding energy of helium and all available neutrons are processed into nuclei.  However, if the dark nucleosynthesis interactions freeze out at temperatures close to, or even a factor of a few below the binding energy, then the dominant asymmetry will remain tied up in the $\pi$ and $\rho$ nucleons.  Thus, already from \Eq{eq:DNscheme}, it is clear that the final asymmetry carried in dark nuclei may vary greater from being a tiny fraction up to the dominant component, depending precisely on when the dark nucleosynthesis interactions freeze out.

In order to study this scenario quantitatively, it is necessary to solve the Boltzmann equations.  In total there are six equations, one for the each baryon and anti-baryon out of each nucleon $\pi$ and $\rho$ and the nuclei $D$.  These equations may be found directly from the Boltzmann equations of \Eq{eq:Boltzmann} by dressing these equations with a label for whether each species carries positive or negative dark baryon number.  In this instance it is crucial to ensure that baryon number is conserved in each interaction, i.e. $Y_\pi^2 \to Y_{\pi^B} Y_{\pi^{\overline{B}}}$ etc.  For the $\pi$ and $\rho$ carrying positive dark baryon number, we have
\begin{eqnarray}
\frac{dY_{\pi^B}}{dx} & = & - \lambda \bigg[R_\pi \left(Y_{\pi^B} Y_{\pi^{\overline{B}}} - {Y_{\pi^B}^{eq}}  {Y_{\pi^{\overline{B}}}^{eq}} \right) +R_N \left(Y_{\pi^B} Y_{D^{\overline{B}}} - \frac{Y_{\rho^{\overline{B}}}}{Y_{\rho^{\overline{B}}}^{eq}} Y_{\pi^B}^{eq} Y_{D^{\overline{B}}}^{eq} \right) \\
&&-  R_N \left(Y_{\rho^{\overline{B}}} Y_{D^B} - \frac{Y_{\pi^B}}{Y_{\pi^B}^{eq}} Y_{\rho^{\overline{B}}}^{eq} Y_{D^B}^{eq} \right)+  R_N f(x) \left(Y_{\pi^B} Y_{\rho^B} - \frac{Y_{D^B}}{Y_{D^B}^{eq}} Y_{\pi^B}^{eq} Y_{\rho^B}^{eq} \right)  \bigg] \,, \nonumber\\
\frac{dY_{\rho^B}}{dx} & = & - \lambda \bigg[R_\rho \left(Y_{\rho^B} Y_{\rho^{\overline{B}}} - {Y_{\rho^B}^{eq}} {Y_{\rho^{\overline{B}}}^{eq}} \right) + R_N \left(Y_{\rho^B} Y_{D^{\overline{B}}} - \frac{Y_{\pi^{\overline{B}}}}{Y_{\pi^{\overline{B}}}^{eq}} Y_{\rho^B}^{eq} Y_{D^{\overline{B}}}^{eq} \right)\, \\
&&-  R_N \left(Y_{\pi^{\overline{B}}} Y_{D^B} - \frac{Y_{\rho^B}}{Y_{\rho^B}^{eq}} Y_{\pi^{\overline{B}}}^{eq} Y_{D^B}^{eq} \right)+  R_N f(x) \left(Y_{\pi^B} Y_{\rho^B} - \frac{Y_{D^B}}{Y_{D^B}^{eq}} Y_{\pi^B}^{eq} Y_{\rho^B}^{eq} \right)  \bigg] \,, \nonumber\\
\label{eq:Boltzmannasymm}
\end{eqnarray}
and for the dark nucleus
\begin{eqnarray}
\frac{dY_{D^B}}{dx} & = & - \lambda \bigg[R_D \left(Y_{D^B} Y_{D^{\overline{B}}} - {Y_{D^B}^{eq}} {Y_{D^{\overline{B}}}^{eq}} \right) -R_N f(x)  \left(Y_{\pi^B} Y_{\rho^B} - \frac{Y_{D^B}}{Y_{D^B}^{eq}} Y_{\pi^B}^{eq} Y_{\rho^B}^{eq} \right)  \nonumber  \\
&&  +R_N \left( \left(Y_{\pi^{\overline{B}}} + Y_{\rho^{\overline{B}}} \right) Y_{D^B} - \left(\frac{Y_{\rho^B}}{Y_{\rho^B}^{eq}} Y_{\pi^{\overline{B}}}^{eq}+ \frac{Y_{\pi^B}}{Y_{\pi^B}^{eq}} Y_{\rho^{\overline{B}}}^{eq} \right) Y_{D^B}^{eq} \right) \bigg] \,. \nonumber
\label{eq:BoltzmannasymmD}
\end{eqnarray}
For the species carrying anti-baryon number, the equations are identical with the exception of the replacement $B \leftrightarrow \overline{B}$.  Considering all six Boltzmann equations and taking the sum $Y_B = Y_{\pi^B}+Y_{\rho^B}+2 Y_{D^B}$ and then by taking the difference $Y_\eta = Y_B - Y_{\overline{B}}$, it is also clear that the dark asymmetry is constant $dY_\eta/dx = 0$, as expected.

In \Fig{fig:thermalasymm}, we show the evolution of the DM abundances in the presence of an asymmetry where we have set the chemical potential in order to generate the observed DM abundance in each case.  As with the symmetric case, dark nuclear capture and dark nucleosynthesis may significantly alter the relic abundance of both the nucleons and the nuclei.  In particular, in the presence of a large dark nucleosynthesis cross section all of the dark $\pi$-mesons may be processed into dark nuclei, leaving only the dark $\rho$-mesons and dark nuclei as the dominant constituents.  As there are three $\rho$ degrees of freedom for every  $\pi$ degree of freedom, once all of the pions are processed into dark nuclei some dark $\rho$ mesons remain.  If they had equal numbers of degrees of freedom, it would be possible for all of the dark nucleons to be processed, leaving only dark nuclei.  This picture is in some ways familiar from the SM where most of the neutrons are processed into nuclei during Big Bang nucleosynthesis, leaving only protons and nuclei.

\section{Indirect Detection Signatures}
\label{sec:indirect}
We will first depart from committing to the specific model of \Sec{sec:model} and instead consider the indirect detection possibilities of dark nucleosynthesis broadly.  In generic scenarios, dark nucleosynthesis may occur via processes such as $n_{n,a}+n_{n,b} \to N_{D,c} + X$ where $n_n$ is a dark nucleon, $N_D$ is a dark nucleus and $X$ is some other state.  If $X$ is a SM state, or if it may decay to SM states, then dark nucleosynthesis occurring presently in DM halos may be observable through the contribution of $X$ to the cosmic ray spectrum.  Considering a particular SM final state $SM$, the spectrum generated in dark nucleosynthesis may be determined from
\be
\label{eq:dphi}
\frac{d^2 \Phi}{d \Omega dE_\gamma} =\frac{1}{8 \pi}  \frac{1}{2 \beta \gamma}\, \zeta\, J(\theta) \int^{E_{SM}/\gamma (1-\beta)}_{E_{SM}/\gamma (1+\beta)} \frac{d\widetilde{E}_{SM}}{\widetilde{E}_{SM}}\frac{dN}{d\widetilde{E}_{SM}} \bigg|_X\,,
\ee
where $J(\theta)$ is the line-of-sight integral over the DM density-squared and $dN/dE_{SM} |_X$ is the spectrum of SM states obtained from $X$ in the rest frame of $X$, either from $X$ directly or from its decays.  In Eq.~(\ref{eq:dphi}), $\gamma$ and $\beta$ are Lorentz factors associated with the fact that $X$ is typically produced with non-zero speed and the integral accommodates the modification of the rest-frame spectrum due to the boosting.  Specifically, for the process $n_{n,a}+n_{n,b} \to N_{D,c} + X$ these factors are given by
\be
\gamma = \frac{(M_a+M_b)^2-M_c^2+M_X^2}{2 (M_a+M_b) M_X}~, \qquad \beta = \sqrt{1-\frac{1}{\gamma^2}}~.
\label{eq:boost}
\ee
$\zeta$ is a factor which is equivalent to $\zeta_{Ann} =2 \langle \sigma v \rangle/M_{DM}^2$ in the case of DM annihilation where the extra factor of $2$ arises as two $X$ states are produced.  In the general case including annihilations and dark nucleosynthesis, this is modified to
\be
\zeta = \kappa_A \sum_{a,b,c} \frac{f_a f_b}{M_a M_b} \langle \sigma v \rangle (n_{n,a}+n_{n,b} \to N_{D,c} + X)~,
\label{eq:zeta}
\ee
where $\kappa_A=1$ for dark nucleosynthesis instead of the usual $\kappa_A=2$ for annihilation. $f_a$ is the fraction of the DM energy density made up by species $a$ and $\langle \sigma v \rangle$ is the thermally-, and spin-averaged cross section and velocity.

If the DM abundance is symmetric, then in general one would also  expect nucleon annihilation signatures from processes such as $n_{n,a}+\overline{n}_{n,a} \to X + X$ and also nuclei annihilation process $N_{D,a}+\overline{N}_{D,a} \to X + X$.  In addition, there could be dark nuclear capture signatures $\overline{n}_{n,a}+N_{D,b} \to n_{n,c} + X$.  If the nucleon mass is $M_n$ the nucleus mass is $M_N = 2 M_n - B_D$ where $B_D\ll M_n$ is the nuclear binding energy.  This provides the main `smoking gun' signature of dark nucleosynthesis which is that in dark nucleosynthesis the energy carried away by $X$ is $E_X \sim \mathcal{O} (B_D \ll M_n)$, however in annihilation or dark nuclear capture the energy carried away is $E_X \sim \mathcal{O} (M_n)$. Thus, if an excess of gamma rays were observed which may be attributed to dark nucleosynthesis (annihilation or capture), then an excess due to the annihilation or capture (dark nucleosynthesis) should also be observable at higher (lower) energies with exactly the same spatial morphology.  Whether or not the other excess is observable depends on both the typical energy scales, and model parameters such as the relative cross sections for dark nucleosynthesis and annihilation.

If the DM abundance is asymmetric, then we are also led to a novel feature of dark nucleosynthesis: in asymmetric DM scenarios it is typically assumed that indirect signatures of DM annihilation cannot be accommodated unless some symmetric DM component is present in the halo.  However, in the case of dark nucleosynthesis if the DM abundance is completely asymmetric then indirect signatures of dark nucleosynthesis are possible and this leads to a novel, and well-motivated, mechanism for generating indirect detection signatures from asymmetric DM.  Specifically, dark baryon number may be conserved in the reaction $n_{n,a}+n_{n,b} \to N_{D,c} + X$, allowing for indirect signatures from asymmetric DM without the need for a symmetric component.

\begin{figure}[t]
  \centering
 \includegraphics[height=0.45\textwidth]{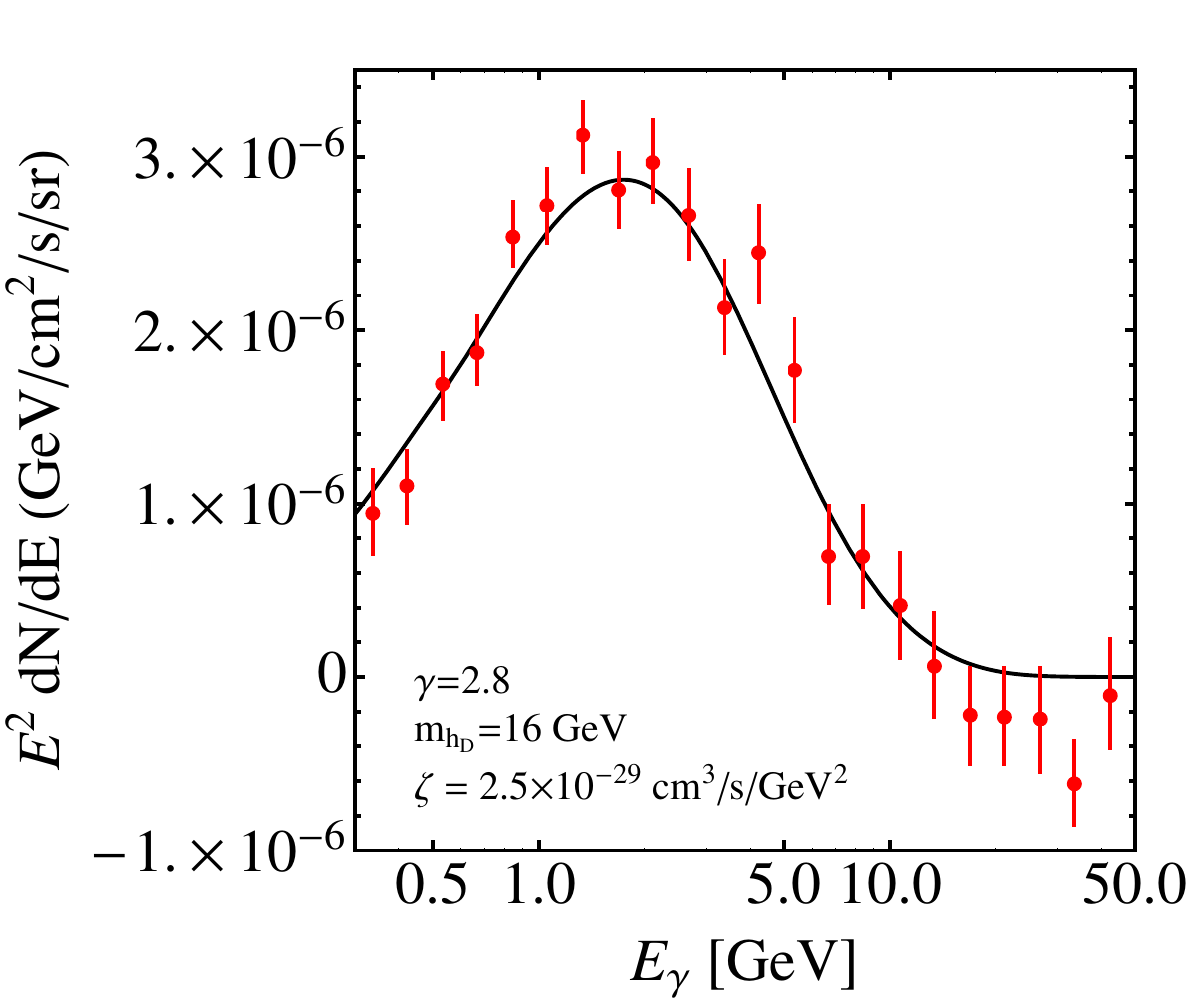}
  \caption{DM parameters which allow an interpretation of the galactic center gamma ray excess.  The red data points show the excess extracted in Ref.~\cite{Daylan:2014rsa} and the black line is the spectrum from boosted $h_D$ decays.  The possible realization of these parameters in specific models is discussed in the text.}
  \label{fig:galcent}
\end{figure}

\subsection{Galactic Center Gamma Ray Excess}
\label{sec:galcentex}
Having discussed the broad indirect detection features of dark nucleosynthesis, we will now show the utility of this process by entertaining the possibility that the gamma ray excess at the galactic center is due to DM \cite{Goodenough:2009gk,Hooper:2010mq,Hooper:2011ti,Abazajian:2012pn,Hooper:2013rwa,Gordon:2013vta,Huang:2013pda,Abazajian:2014fta,Daylan:2014rsa}, specifically considering an interpretation in terms of dark nucleosynthesis or capture.\footnote{It should be noted that plausible  interpretations based on SM physics have also been suggested \cite{Carlson:2014cwa,Petrovic:2014uda}, and thus we use this DM hint as an interesting scenario with which to demonstrate the possible indirect detection signatures of dark nuclei, but not as the main motivation for this work.}  With regard to dark sector-SM interactions, we envisage the model of \Sec{sec:model} in which $X$ is a light singlet scalar with a small mixing with the SM Higgs boson, identified previously as a dark Higgs $h_D$.  For masses $M_{h_D} > 2 m_b$ and $M_{h_D} < 2 m_W$ the dominant decay mode of the dark Higgs will be to a pair of $b$-quarks.

To fit the spectrum, we employ the prompt gamma ray spectrum from $b$-quarks obtained in Ref.~\cite{Cirelli:2010xx}.\footnote{We do not include final-state effects such as bremsstrahlung for this analysis, but note that these effects may lead to small quantitative changes to the spectrum \cite{Cirelli:2013mqa}.}  We calculate the $J$-factor for the best-fit NFW \cite{Navarro:1995iw} profile of Ref.~\cite{Daylan:2014rsa} with scale-radius $r_S = 20$ kpc, and choose the overall density parameter such that the local DM density at $8.5$kpc is $0.3\text{ GeV cm}^{-3}$.  We also choose the NFW profile parameter $\gamma = 1.26$.  The spectrum of \cite{Daylan:2014rsa} is normalized to the spectrum at $\theta = 5^{\circ}$ and we find $J(5^{\circ}) = 6.2 \times 10^{23} \text{ GeV}^2 \text{cm}^{-5}$.  There are many parameter choices which may give a reasonable fit to the data and in \Fig{fig:galcent} we show one parameter choice allowing a good fit to the data where $M_{h_D} = 16$ GeV and the dark Higgs is produced at a boost of $\gamma = 2.8$.  This explanation requires $\zeta = 2.5 \times 10^{-29} \text{cm}^3 \text{ s}^{-1} \text{ GeV}^{-2} $, providing a target for an interpretation of this excess.  However, it is worth emphasizing that all of these numbers may change with different choices of local DM density, halo profiles, different template fitting procedures to extract the gamma ray excess, and also with different SM final states, thus it should be kept in mind that the required parameters are a good qualitative guide but are subject to a number of uncertainties.

\subsubsection{A Dark Nuclear Capture Interpretation}
\label{sec:capture}
If the DM is symmetric, then it is possible for indirect detection signals to arise in a number of ways.  The first, and very well known, possibility is for DM annihilations.  In this context, the gamma ray excess in the galactic center may be easily accommodated through the annihilation of nucleons, or nuclei, of mass $\sim 45$ GeV into pairs of dark Higgs bosons which eventually decay to $\overline{b}b$ pairs.  As this scenario is very well known we will not dwell on it any further.

\begin{figure}[]
  \centering
 \includegraphics[height=0.2\textwidth]{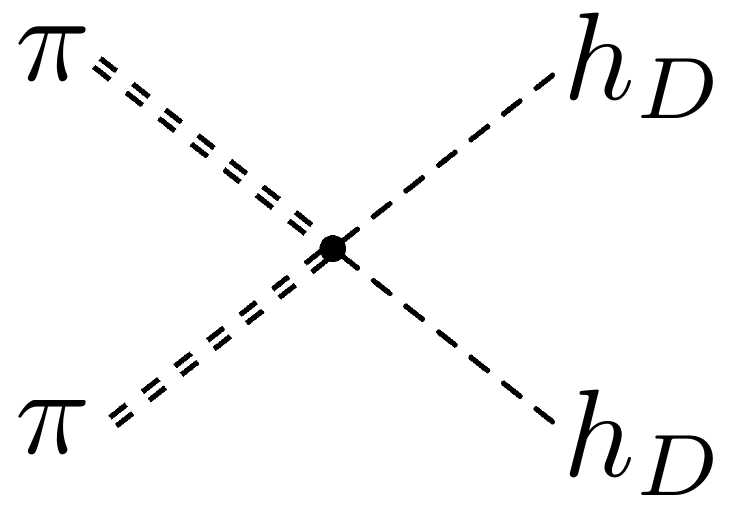} \qquad \includegraphics[height=0.2\textwidth]{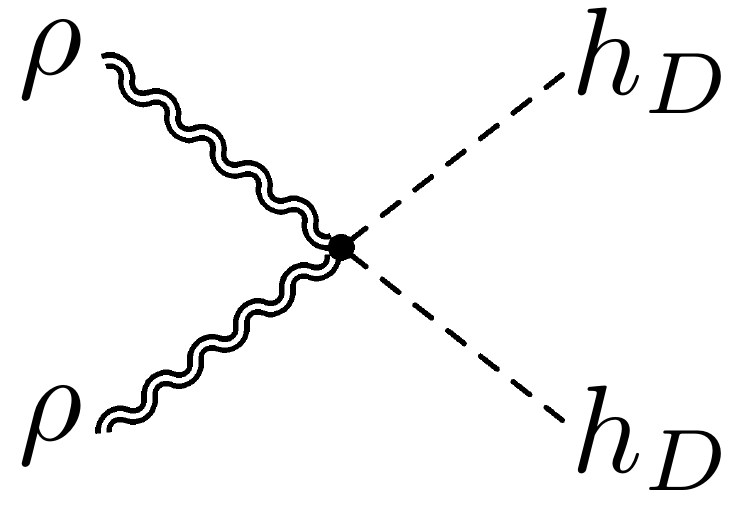} \qquad\includegraphics[height=0.2\textwidth]{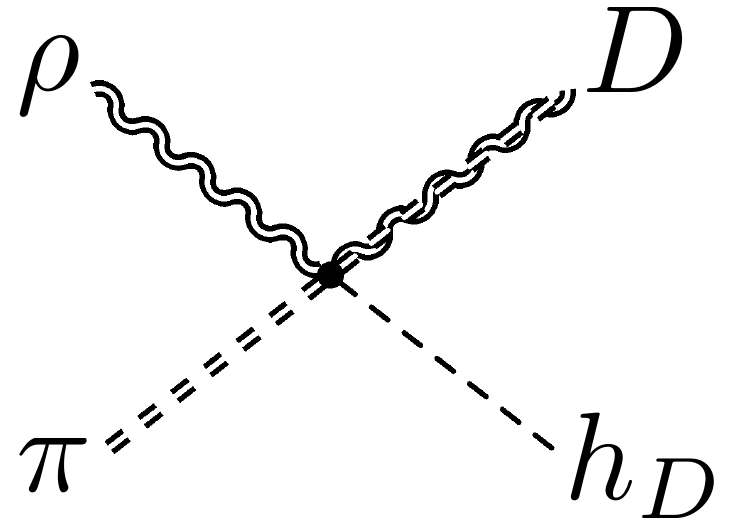}
  \caption{Annihilation and dark nucleosynthesis processes leading to indirect detection signatures of symmetric DM.  Rearrangements of the final diagram involving dark nuclear capture $D+(\pi, \rho) \to h_D + (\rho,\pi)$ are also possible.}
  \label{fig:symmdiag}
\end{figure}

Another scenario, which has not been considered previously, is relevant if a symmetric component of dark nuclei is regenerated in the early Universe as in \Fig{fig:thermal}.  In this case, it is possible for indirect detection signals to arise through dark nuclear capture processes such as $\pi^a +D^b \to \rho^c +h_D$, followed by $h_D \to \overline{b}b$.  In this case the dark nucleosynthesis process is critically important, both for regenerating the dark nuclei in the early Universe and also for the capture which leads to potential signals.  Some of the indirect detection signatures possible in this scenario are depicted in \Fig{fig:symmdiag} and their associated energy scales are given in \Tab{tab:symmsig}.

\begin{figure}[t]
  \centering
 \includegraphics[height=0.45\textwidth]{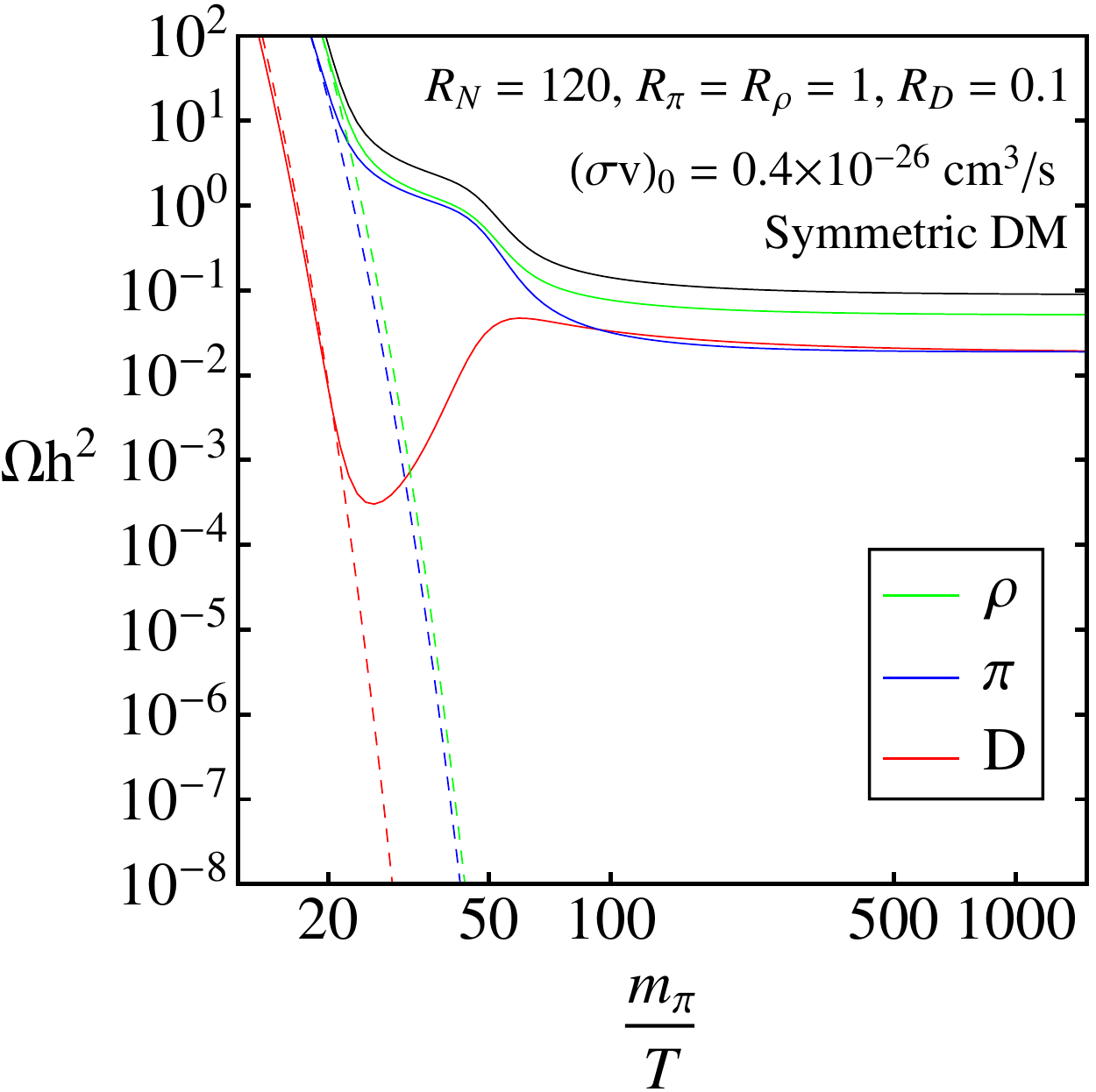} \qquad  \includegraphics[height=0.46\textwidth]{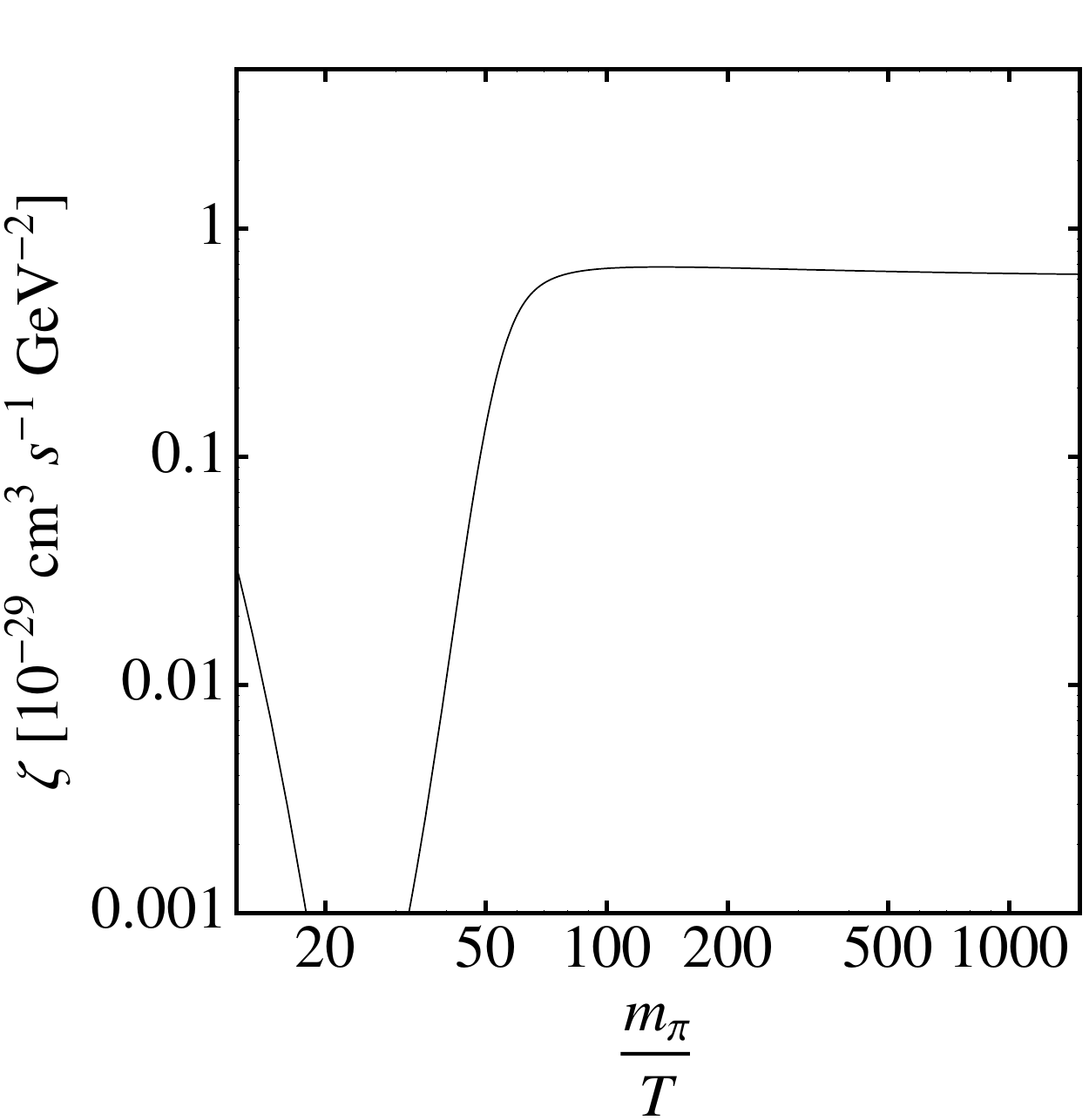}
  \caption{Evolution of cosmological relic DM densities (left panel) and the $\zeta$-factor for indirect detection defined in \Eq{eq:zeta} (right panel).  This is a particular parameter choice which gives rise to the galactic center gamma ray excess from dark nucleus destruction processes occurring in the center of the galaxy.  The full solutions are shown as solid lines and the equilibrium values as dashed lines.}
  \label{fig:indirectsymm}
\end{figure}

In \Fig{fig:indirectsymm}, we consider a scenario that is motivated by the model of \Sec{sec:model}.  The nucleon masses are both taken to be $M_\pi = M_\rho = 40$ GeV.  The dark Higgs mass is $M_{h_D} = 16$ GeV and we allow for dark nucleosynthesis only at the kinematic threshold such that $\delta = M_{h_D}/M_\pi$.\footnote{This binding energy is quite large, of $\mathcal{O} (40\%)$ the nucleon mass and may thus not lie strictly within the confines of the $\SU(2)$ model, however in this section we wish to explore general possibilities for dark nucleosynthesis and choose this binding such that the on-threshold Boltzmann equations of \Sec{sec:freeze} may be used.}  The masses are chosen such that in dark nuclear capture the dark Higgs bosons are produced with a boost factor of $2.8$, as desired. 

In \Fig{fig:indirectsymm}, we show the additional parameters of the model.  In the left panel it is shown that the observed relic density may be achieved for these parameters, and in the right panel the $\zeta$-factor for indirect detection is shown.  From this we see that the $\zeta$-factor is too low by approximately a factor of four, however (as argued in Ref.~\cite{Boehm:2014bia}, for example) specific choices about the form of the halo lead to the required value of $\zeta = 2.5 \times 10^{-29} \text{cm}^3 \text{s}^{-1} \text{GeV}^2$ and thus a different choice of DM halo profile, particularly in the center of the galaxy, could account for this additional factor of four.  Other final states could also be considered, which may accommodate smaller cross sections.

Thus we see that nuclear processes in a symmetric dark sector may lead to a novel cosmology and a novel interpretation of the galactic center gamma ray excess.  Furthermore in this scenario additional, but greatly subdominant, nucleon and nucleus annihilation signatures would also be present with greater boost factors ($\mathcal{O}(3.8)$) however the fluxes are small enough, and the boost factors similar enough, that this would only moderately change the spectrum.  For these parameters, dark nucleosynthesis is at threshold, and thus indirect signatures of dark nucleosynthesis would not be expected.

\subsection{Indirect Signals of Asymmetric Dark Nucleosynthesis}\label{sec:indirectnucleo}
An interesting feature which is raised by (but not restricted to) dark nucleosynthesis is the possibility of indirect signals of purely asymmetric dark matter.  In single-component models of purely asymmetric dark matter it has long been known that indirect detection signals are not possible as annihilation of thermal relics is not compatible with a conserved global $\U(1)$ symmetry in the dark sector.  Some authors have considered annihilations involving a small relic, or regenerated, symmetric DM component, but this is not possible in strictly asymmetric DM scenarios \cite{Buckley:2011ye,Cirelli:2011ac,Tulin:2012re,Okada:2012rm,Hardy:2014dea}. 

\begin{figure}[]
  \centering
 \includegraphics[height=0.2\textwidth]{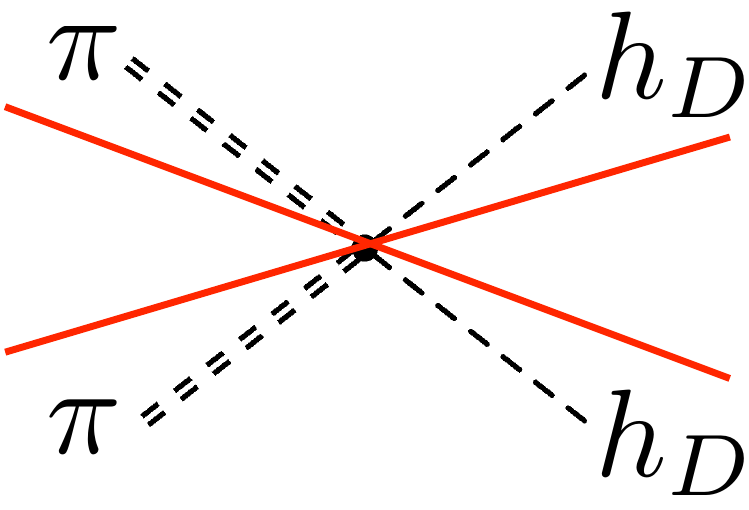} \qquad \includegraphics[height=0.2\textwidth]{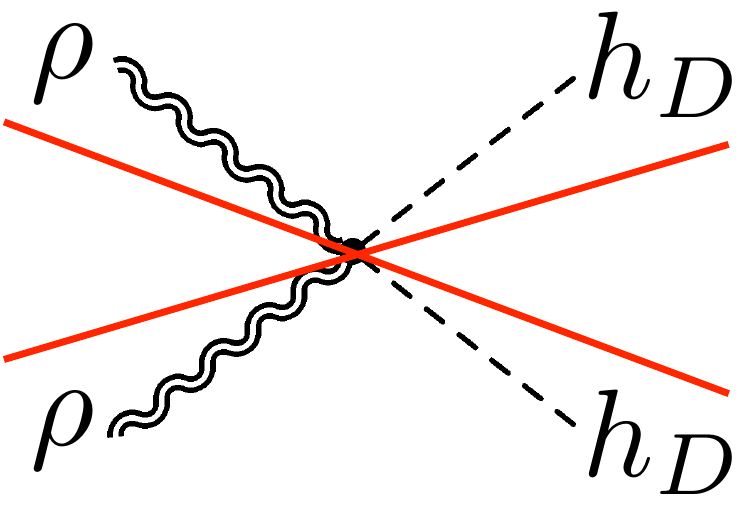} \qquad\includegraphics[height=0.2\textwidth]{figures/SemiAnn.pdf}
  \caption{Indirect detection signatures of asymmetric DM.  Rearrangements of the final diagram involving dark nuclear destruction $D+\pi, \rho \to h_D + \rho,\pi$ are not possible due to dark baryon number conservation. The diagrams with crosses are forbidden in asymmetric DM scenarios, however dark nucleosynthesis is still possible.}
  \label{fig:indirectasymmdiags}
\end{figure}

\begin{table}[]
\centering
\begin{tabular}{ c | c | c | c | c | c}
Signature & Collider & Direct Detection & Annihilation & Nucleosynthesis & Capture\\
\hline \hline
Sym-DM & $M,2M$ & $M,2M$ & $M,2M$ & $E\ll M$ & $M$ \\
Asym-DM & $M,2M$ & $M,2M$ & --- & $E\ll M$ & ---
\end{tabular}
\caption{Typical energy scales associated with symmetric and asymmetric DM signatures, where the mass $M$ denotes the typical nucleon mass.  Unlike symmetric DM, annihilation signals are absent for purely asymmetric DM, however indirect signals may still arise for dark nucleosynthesis in this model, or more general multi-component asymmetric DM models.}
\label{tab:symmsig}
\end{table}

However, if the dark sector involves more than one stable state it {\it is} possible to have indirect detection signals for purely asymmetric dark matter while conserving the global DM symmetry.  A classic analogue of this arises in the SM where the nucleosynthesis process $n+p \to D + \gamma$ conserves baryon number.  Following this analogy, in ADM scenarios such processes may still be observable in the current epoch, raising the intriguing possibility of indirect detection signals from a fully asymmetric dark sector.  In the specific model considered here, the analogous process is $\pi^B+\rho^B \to D^B+h_D$.  In this section, we will study possible signals from this process, however it should be emphasized that these signals are possible in a great variety of asymmetric DM models and are not restricted to nuclear or composite DM.  The full range of possibilities is deserving of a dedicated study and here we just consider a variant of the dark nuclear model of \Sec{sec:model}.  Indirect detection signatures possible in this scenario are depicted in \Fig{fig:indirectasymmdiags} and their associated energy scales are given in \Tab{tab:symmsig}.

\subsubsection{Galactic Signals of Asymmetric Dark Nucleosynthesis}\label{sec:galexcess}

For the sake of simplicity, we will consider the same asymmetric DM model of \Sec{sec:model} where the only states are the dark baryon number carrying states $\pi^B$, $\rho^B$, and the dark nucleus $D^B$ which carries dark baryon number two.  Attempting to explain the gamma ray excess as in Sec.~\ref{sec:capture}, we choose the same parameters as before, with $m_{h_D} = 16$ GeV and boost factor $\gamma=2.8$.  Assuming heavy DM, $M_\pi = M_\rho = 250$ GeV, then the correct boost factor may be achieved with a nuclear binding energy fraction of $\delta \approx 0.2$.

\begin{figure}[]
  \centering
 \includegraphics[height=0.45\textwidth]{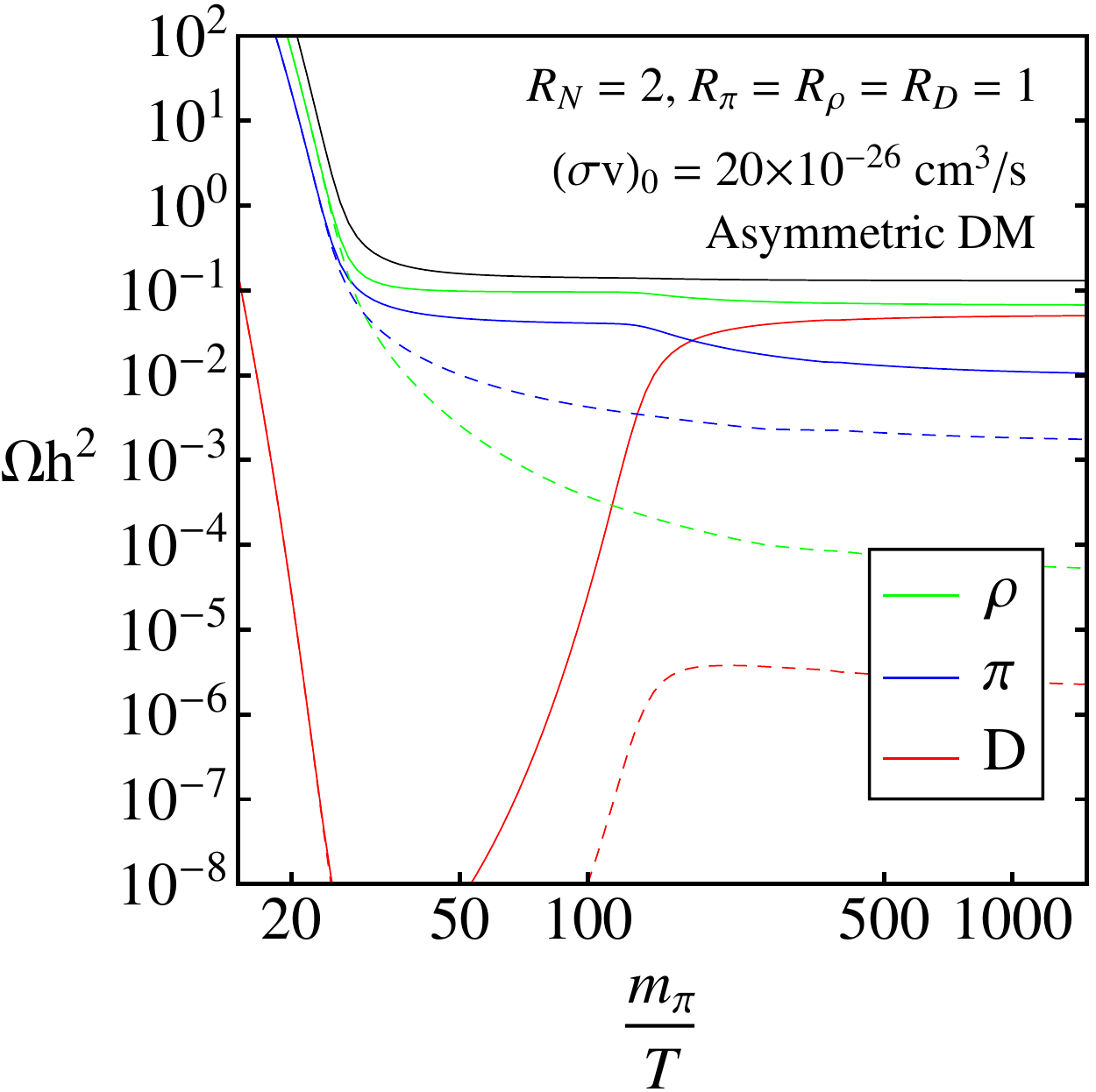} \qquad  \includegraphics[height=0.46\textwidth]{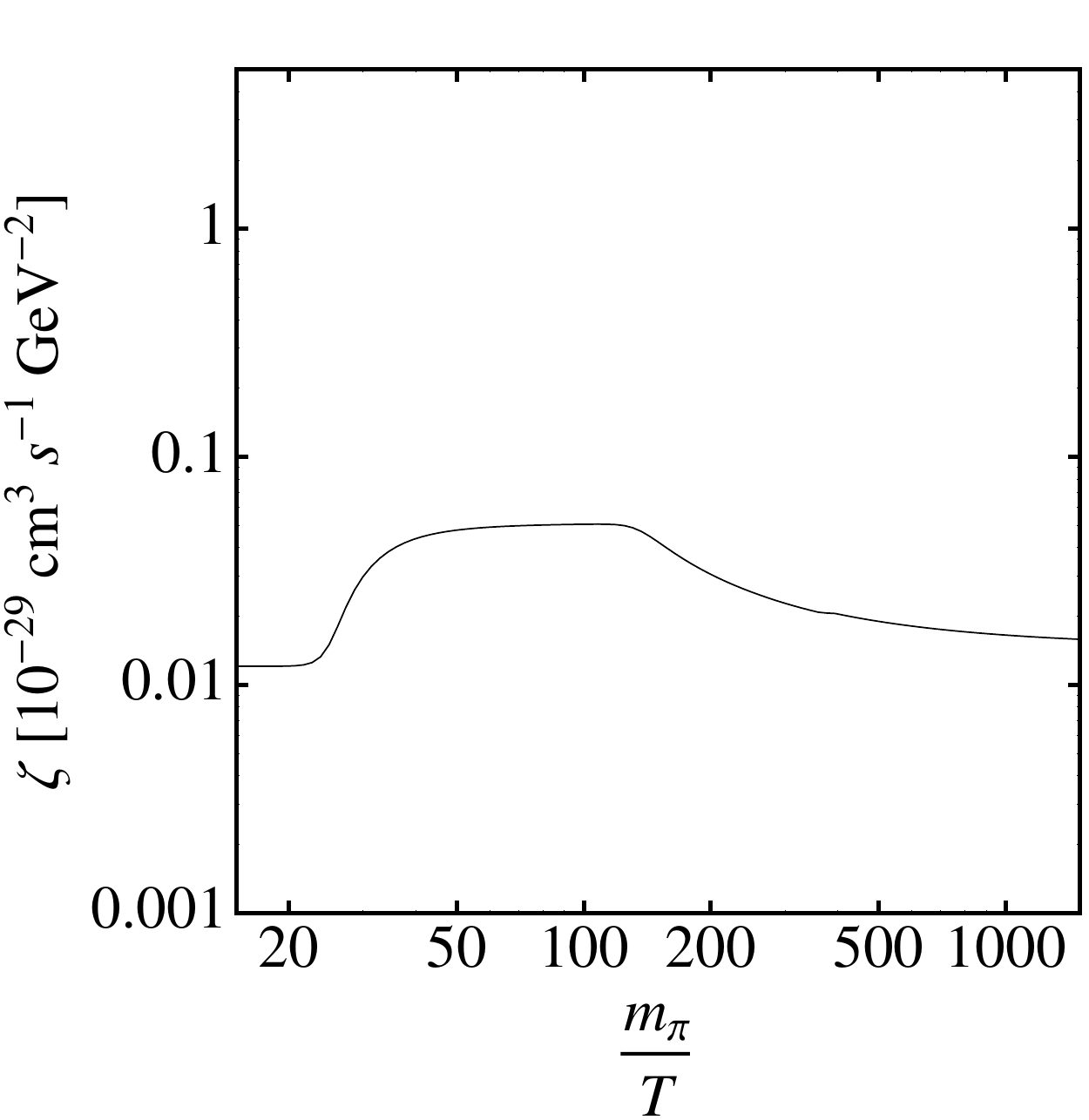}
  \caption{Evolution of cosmological relic DM densities (left panel) and the $\zeta$-factor for indirect detection defined in \Eq{eq:zeta} (right panel) for asymmetric DM.  This is a particular parameter choice aimed at explaining the galactic center gamma ray excess from dark nucleus destruction processes occurring in the center of the galaxy.  In the left panel the dark baryons are shown as solid lines and dark anti-baryons as dashed lines.  The right panel demonstrates that within this model for these chosen parameters an explanation of the galactic center excess based on dark nucleosynthesis is unlikely.}
  \label{fig:indirectasymm}
\end{figure}

In \Fig{fig:indirectasymm}, we show the the cosmological evolution of an asymmetric DM scenario for this specific choice of parameters.  The nucleosynthesis cross section has been taken large enough that the majority of $\pi^B$ have been processed into nuclei by the time the evolution stabilizes.  In the right hand plot, we show the $\zeta$-factor relevant for the galactic gamma ray excess.  For this case, we see that this factor is too small by two orders of magnitude.  This is due to a number of factors.  First, the total energy released in dark nucleosynthesis is the binding energy which is $B_D = \delta \, M_\pi \ll M_\pi$. To boost a $16$ GeV dark Higgs by a sufficient amount while keeping the binding energy fraction small enough to identify $D^B$ as a bound state of two nucleons requires relatively heavy DM, $M_\pi \gg m_{h_D}$.  Since the number density is inversely proportional to the square of this number, this significantly suppresses the signal.  Second, for the asymmetric DM scenario, the dark nucleosynthesis cross section may not be taken arbitrarily large as then all of the available $\pi^B$ mesons will be processed into nuclei in the early Universe and too few $\pi^B$ will remain in the current epoch to nucleosynthesize and generate the observed gamma ray excess.

Overall, it seems that within the confines of this simplest version of a dark nuclei model, an explanation of the galactic center gamma ray excess appears difficult for an asymmetric DM scenario with dark nucleosynthesis.  It should be emphasized that this is only within this specific model and an asymmetric DM interpretation is not precluded on general grounds. It would be interesting to explore this scenario by considering other halo profiles and/or SM final states. Indeed, this example demonstrates that dark nucleosynthesis allows for indirect signals of asymmetric DM even in the absence of any symmetric DM component.

\subsubsection{Asymmetric Dark Nucleosynthesis and Solar Capture}\label{sec:solar}
If DM scatters on SM nucleons, it may become captured in astrophysical hosts, such as planets, stars such as the Sun \cite{Spergel:1984re,Press:1985ug,Silk:1985ax,Srednicki:1986vj,Gould:1987ju,Gould:1987ir,Gould:1987ww,Steigman:1997vs,Bottino:2002pd}, neutron stars and white dwarfs \cite{Moskalenko:2007ak,Bertone:2007ae,McCullough:2010ai}.  In the context of asymmetric DM, it is assumed that because of the lack of DM annihilations, the abundance of asymmetric DM will gradually build up in these objects and eventually alter their properties \cite{Frandsen:2010yj,Kouvaris:2010jy,McDermott:2011jp,Kouvaris:2011fi,Iocco:2012wk,Lopes:2012af,Bell:2013xk}, in some cases quite spectacularly through modifications of helioseismology or even the premature gravitational collapse of neutron stars.  However, if the possibility of dark nucleosynthesis is introduced, the phenomenology of asymmetric DM capture may be altered radically.  We leave a full quantitative study to future work and only discuss potential qualitative signatures here.

\begin{figure}[]
  \centering
 \includegraphics[height=0.3\textwidth]{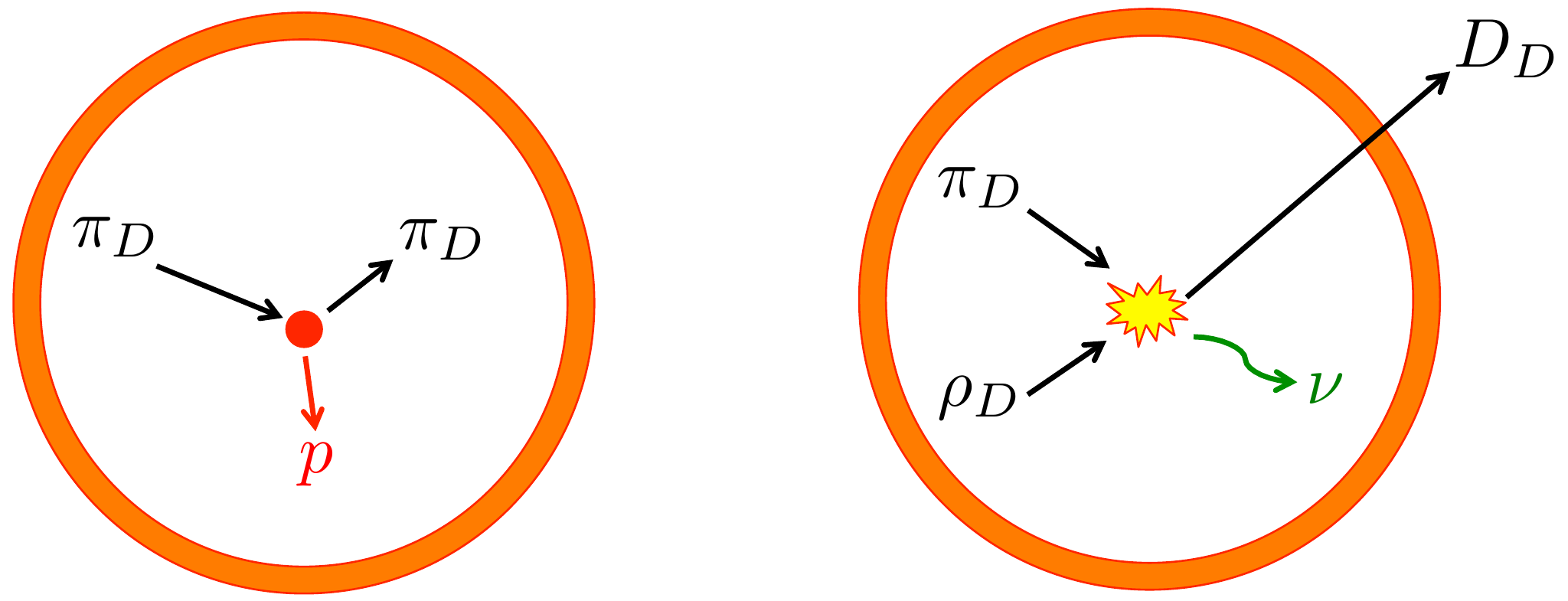}
  \caption{Capture of asymmetric DM in astrophysical bodies such as planets, the Sun, white dwarfs, and neutron stars (left panel).  Dark nucleosynthesis in these astrophysical bodies is catalyzed by the enhanced density of DM (right panel).  Dark nucleosynthesis may lead to observable signatures if the end-products produce neutrinos either through decay or rescattering.  Even if the binding energy fraction is small, the produced dark nucleus may be ejected from the astrophysical body because the resulting semi-relativistic velocity of the dark nucleus would typically be greater than the escape velocity.  This may drastically alter the phenomenology of asymmetric DM capture in comparison to standard asymmetric DM models, and the ejected dark nuclei could be searched for in new laboratory experiments.}
  \label{fig:solar}
\end{figure}

If they scatter on SM nucleons, dark nucleons and nuclei would steadily build up within a star as in  standard DM models.  However, unlike standard asymmetric DM scenarios, dark nucleosynthesis would also occur within the star due to the increasing density of DM.  In this case, dark nucleosynthesis may lead to observable indirect detection signatures from the Earth or the Sun if the neutral dark nucleosynthesis final states include SM particles that can subsequently produce observable neutrinos through decay or rescattering, as depicted in \Fig{fig:solar}.  This is not possible for standard asymmetric DM candidates.

Another interesting feature of dark nucleosynthesis is that even for very small binding energies, the produced dark nuclei may have a semi-relativistic velocity allowing it to escape the astrophysical host.  In general, this occurs for $\beta > \beta_{\text{Escape}}$.  For dark nucleosynthesis with binding fraction $\delta$ and with a massless neutral final state particle, the outgoing speed of the nucleus is $\beta \approx \delta/2$, thus for a binding energy fraction
$\delta \gtrsim 0.01$,
the dark nucleus would be ejected from the Sun by dark nucleosynthesis.  Dark matter ejection due to dark nucleosynthesis could thus have a significant effect as the usual build up of asymmetric DM may be obstructed.  For the Sun, the expected modifications of helioseismology may be reduced.  For more compact objects, the build up of a large DM component would be slowed, or even avoided, due to the steady ejection of DM from the star.  Furthermore, it may be possible to search for these ejected dark nuclei in Earth-based laboratory experiments by searching for neutral-current scattering events in low-background detectors where the scattering energy is at an energy scale of $E \sim \delta M_{DM}$ and the incoming dark nucleus points towards the Sun or the center of the Earth.  This signature would motivate similar searches as recently proposed in \cite{Huang:2013xfa,Agashe:2014yua}, however at potentially lower energy scales.

There is also a very pleasing synergy between DM and the visible sector in this case as the capture of asymmetric DM in stars leads to the dark nucleons being processed into dark nuclei, in a tenuous analogy with the processes which occur in the visible sector.  If there are additional dark nuclei with larger dark baryon number, further dark nucleosynthesis may also occur, processing the dark nucleons into more massive dark nuclei. In essence, the star would lead to a co-located dark protostar, burning dark nucleons into dark nuclei.  All of these features require a detailed study for a full exploration of the capture and ejection processes, and a dedicated study of the experimental requirements for detecting the ejected dark nuclei is also required. However, our brief discussion is suggestive of a very rich and novel phenomenology which could lead to experimental signatures significantly different from those expected of standard DM candidates.

\section{Conclusions}\label{sec:conclusions}
To ensure that possible experimental signatures of DM are not missed, it is crucial to consider the broad scope of possible realizations of DM, in addition to the more well-studied DM candidates.  From a theoretical perspective, the possibility of dark nuclear physics is well motivated.  In fact, in the two strongly-coupled theories for which nuclear states have been studied, the SM and two-color two-flavor QCD, nuclei are seen to exist. 
For QCD, nuclei have also been shown to occur for heavier-than-physical quark masses \cite{Beane:2012vq,Detmold:2012eu,Yamazaki:2012hi}.
 As far as quantitatively studied strongly-coupled composites are concerned, this hints towards the ubiquity of nuclei.  Thus, if DM consists of composites of a strongly coupled gauge sector, then it is very possible that there is an entire dark nuclear sector.

In this work, motivated by the lattice results to be presented in a companion paper, and by analogy with the SM, some aspects of dark nuclear phenomenology have been explored.  For symmetric and asymmetric DM, it is possible that the abundance may be composed of a range of admixtures of dark nucleons and dark nuclei.  New indirect detection possibilities have been found, and an illustrative explanation of the galactic center gamma ray excess based on dark nuclear capture has been presented.  For asymmetric DM, the consequences of dark nuclei are striking.  Dark nucleosynthesis accommodates indirect detection signatures of asymmetric DM, even for a vanishing symmetric component.  This opens new avenues for asymmetric DM model building.

The phenomenology of DM capture in astrophysical bodies may also be significantly modified.  Not only are indirect detection signals of captured asymmetric DM possible, but dark nucleosynthesis may also radically alter the process of capture.  Even for small binding energy fractions, dark nucleosynthesis may lead to the ejection of asymmetric dark nuclei from stars, suppressing the build up asymmetric DM in these objects.  There is also possibly  an attractive synergy between the dark and visible sectors in which visible stars essentially catalyze the production of dark nuclei.

By touching upon the broad phenomenological features of dark nuclei, important departures from the standard signatures of DM have been demonstrated, particularly for the  scenario of asymmetric DM.  It has also been argued that dark nuclear physics is a well-motivated consideration for the dark sector.  It would be interesting to  map out further possibilities by considering different models, particularly with guidance from lattice field theory methods, which may exhibit different confining gauge symmetries, different global symmetry breaking patterns, different flavor symmetries, and also heavier nuclei.  It would also be interesting to study more broadly the early Universe cosmology, indirect detection, solar capture, and direct detection possibilities.  Our current studies suggest that the general phenomenology of dark nuclei is rich.

\acknowledgments{We thank Timothy Cohen, Patrick Fox, Michael Ramsey-Musolf, Jesse Thaler, Martin Savage, Brian Shuve, Tracey Slatyer, James Unwin, Neal Weiner and the participants of the KITP `Particlegenesis' program for useful conversations. M.M. is grateful for the hospitality of the KITP during the completion of this work and is supported by a Simons Postdoctoral Fellowship, W.D by a US Department of Energy Early Career Research Award DE-SC0010495 and the Solomon Buchsbaum Fund at MIT and AVP by Department of Energy grant DE-FG02-94ER40818.}

\bibliographystyle{JHEP}
\bibliography{PhenoDraft}

\providecommand{\href}[2]{#2}\begingroup\raggedright\begin{thebibliography}{100}

\bibitem{Alves:2010dd}
D.~Spier Moreira~Alves, S.~R. Behbahani, P.~Schuster, and J.~G. Wacker, {\it
  {The Cosmology of Composite Inelastic Dark Matter}},  {\em JHEP} {\bf 1006}
  (2010) 113, [\href{http://xxx.lanl.gov/abs/1003.4729}{{\tt
  arXiv:1003.4729}}].

\bibitem{Behbahani:2010xa}
S.~R. Behbahani, M.~Jankowiak, T.~Rube, and J.~G. Wacker, {\it {Nearly
  Supersymmetric Dark Atoms}},  {\em Adv.High Energy Phys.} {\bf 2011} (2011)
  709492, [\href{http://xxx.lanl.gov/abs/1009.3523}{{\tt arXiv:1009.3523}}].

\bibitem{Kaplan:2011yj}
D.~E. Kaplan, G.~Z. Krnjaic, K.~R. Rehermann, and C.~M. Wells, {\it {Dark
  Atoms: Asymmetry and Direct Detection}},  {\em JCAP} {\bf 1110} (2011) 011,
  [\href{http://xxx.lanl.gov/abs/1105.2073}{{\tt arXiv:1105.2073}}].

\bibitem{Kumar:2011iy}
K.~Kumar, A.~Menon, and T.~M. Tait, {\it {Magnetic Fluffy Dark Matter}},  {\em
  JHEP} {\bf 1202} (2012) 131, [\href{http://xxx.lanl.gov/abs/1111.2336}{{\tt
  arXiv:1111.2336}}].

\bibitem{Khlopov:2011tn}
M.~Y. Khlopov, {\it {Physics of Dark Matter in the Light of Dark Atoms}},  {\em
  Mod.Phys.Lett.} {\bf A26} (2011) 2823--2839,
  [\href{http://xxx.lanl.gov/abs/1111.2838}{{\tt arXiv:1111.2838}}].

\bibitem{Cline:2012is}
J.~M. Cline, Z.~Liu, and W.~Xue, {\it {Millicharged Atomic Dark Matter}},  {\em
  Phys.Rev.} {\bf D85} (2012) 101302,
  [\href{http://xxx.lanl.gov/abs/1201.4858}{{\tt arXiv:1201.4858}}].

\bibitem{CyrRacine:2012fz}
F.-Y. Cyr-Racine and K.~Sigurdson, {\it {Cosmology of atomic dark matter}},
  {\em Phys.Rev.} {\bf D87} (2013), no.~10 103515,
  [\href{http://xxx.lanl.gov/abs/1209.5752}{{\tt arXiv:1209.5752}}].

\bibitem{Fan:2013yva}
J.~Fan, A.~Katz, L.~Randall, and M.~Reece, {\it {Double-Disk Dark Matter}},
  {\em Phys.Dark Univ.} {\bf 2} (2013) 139--156,
  [\href{http://xxx.lanl.gov/abs/1303.1521}{{\tt arXiv:1303.1521}}].

\bibitem{Fan:2013tia}
J.~Fan, A.~Katz, L.~Randall, and M.~Reece, {\it {Dark-Disk Universe}},  {\em
  Phys.Rev.Lett.} {\bf 110} (2013), no.~21 211302,
  [\href{http://xxx.lanl.gov/abs/1303.3271}{{\tt arXiv:1303.3271}}].

\bibitem{McCullough:2013jma}
M.~McCullough and L.~Randall, {\it {Exothermic Double-Disk Dark Matter}},  {\em
  JCAP} {\bf 1310} (2013) 058, [\href{http://xxx.lanl.gov/abs/1307.4095}{{\tt
  arXiv:1307.4095}}].

\bibitem{Cline:2013pca}
J.~M. Cline, Z.~Liu, G.~Moore, and W.~Xue, {\it {Scattering properties of dark
  atoms and molecules}},  {\em Phys.Rev.} {\bf D89} (2014) 043514,
  [\href{http://xxx.lanl.gov/abs/1311.6468}{{\tt arXiv:1311.6468}}].

\bibitem{Belotsky:2014haa}
K.~Belotsky, M.~Khlopov, C.~Kouvaris, and M.~Laletin, {\it {Decaying Dark Atom
  constituents and cosmic positron excess}},  {\em Adv.High Energy Phys.} {\bf
  2014} (2014) 214258, [\href{http://xxx.lanl.gov/abs/1403.1212}{{\tt
  arXiv:1403.1212}}].

\bibitem{Nussinov:1985xr}
S.~Nussinov, {\it {TECHNOCOSMOLOGY: COULD A TECHNIBARYON EXCESS PROVIDE A
  'NATURAL' MISSING MASS CANDIDATE?}},  {\em Phys.Lett.} {\bf B165} (1985) 55.

\bibitem{Barr:1990ca}
S.~M. Barr, R.~S. Chivukula, and E.~Farhi, {\it {Electroweak Fermion Number
  Violation and the Production of Stable Particles in the Early Universe}},
  {\em Phys.Lett.} {\bf B241} (1990) 387--391.

\bibitem{Khlopov:2005ew}
M.~Y. Khlopov, {\it {Composite dark matter from 4th generation}},  {\em Pisma
  Zh.Eksp.Teor.Fiz.} {\bf 83} (2006) 3--6,
  [\href{http://xxx.lanl.gov/abs/astro-ph/0511796}{{\tt astro-ph/0511796}}].

\bibitem{Gudnason:2006ug}
S.~B. Gudnason, C.~Kouvaris, and F.~Sannino, {\it {Towards working technicolor:
  Effective theories and dark matter}},  {\em Phys.Rev.} {\bf D73} (2006)
  115003, [\href{http://xxx.lanl.gov/abs/hep-ph/0603014}{{\tt
  hep-ph/0603014}}].

\bibitem{Gudnason:2006yj}
S.~B. Gudnason, C.~Kouvaris, and F.~Sannino, {\it {Dark Matter from new
  Technicolor Theories}},  {\em Phys.Rev.} {\bf D74} (2006) 095008,
  [\href{http://xxx.lanl.gov/abs/hep-ph/0608055}{{\tt hep-ph/0608055}}].

\bibitem{Khlopov:2008ty}
M.~Y. Khlopov and C.~Kouvaris, {\it {Composite dark matter from a model with
  composite Higgs boson}},  {\em Phys.Rev.} {\bf D78} (2008) 065040,
  [\href{http://xxx.lanl.gov/abs/0806.1191}{{\tt arXiv:0806.1191}}].

\bibitem{Ryttov:2008xe}
T.~A. Ryttov and F.~Sannino, {\it {Ultra Minimal Technicolor and its Dark
  Matter TIMP}},  {\em Phys.Rev.} {\bf D78} (2008) 115010,
  [\href{http://xxx.lanl.gov/abs/0809.0713}{{\tt arXiv:0809.0713}}].

\bibitem{Foadi:2008qv}
R.~Foadi, M.~T. Frandsen, and F.~Sannino, {\it {Technicolor Dark Matter}},
  {\em Phys.Rev.} {\bf D80} (2009) 037702,
  [\href{http://xxx.lanl.gov/abs/0812.3406}{{\tt arXiv:0812.3406}}].

\bibitem{Alves:2009nf}
D.~S. Alves, S.~R. Behbahani, P.~Schuster, and J.~G. Wacker, {\it {Composite
  Inelastic Dark Matter}},  {\em Phys.Lett.} {\bf B692} (2010) 323--326,
  [\href{http://xxx.lanl.gov/abs/0903.3945}{{\tt arXiv:0903.3945}}].

\bibitem{Mardon:2009gw}
J.~Mardon, Y.~Nomura, and J.~Thaler, {\it {Cosmic Signals from the Hidden
  Sector}},  {\em Phys.Rev.} {\bf D80} (2009) 035013,
  [\href{http://xxx.lanl.gov/abs/0905.3749}{{\tt arXiv:0905.3749}}].

\bibitem{Kribs:2009fy}
G.~D. Kribs, T.~S. Roy, J.~Terning, and K.~M. Zurek, {\it {Quirky Composite
  Dark Matter}},  {\em Phys.Rev.} {\bf D81} (2010) 095001,
  [\href{http://xxx.lanl.gov/abs/0909.2034}{{\tt arXiv:0909.2034}}].

\bibitem{Frandsen:2009mi}
M.~T. Frandsen and F.~Sannino, {\it {iTIMP: isotriplet Technicolor Interacting
  Massive Particle as Dark Matter}},  {\em Phys.Rev.} {\bf D81} (2010) 097704,
  [\href{http://xxx.lanl.gov/abs/0911.1570}{{\tt arXiv:0911.1570}}].

\bibitem{Lisanti:2009am}
M.~Lisanti and J.~G. Wacker, {\it {Parity Violation in Composite Inelastic Dark
  Matter Models}},  {\em Phys.Rev.} {\bf D82} (2010) 055023,
  [\href{http://xxx.lanl.gov/abs/0911.4483}{{\tt arXiv:0911.4483}}].

\bibitem{Khlopov:2010pq}
M.~Y. Khlopov, A.~G. Mayorov, and E.~Y. Soldatov, {\it {Composite Dark Matter
  and Puzzles of Dark Matter Searches}},  {\em Int.J.Mod.Phys.} {\bf D19}
  (2010) 1385--1395, [\href{http://xxx.lanl.gov/abs/1003.1144}{{\tt
  arXiv:1003.1144}}].

\bibitem{Belyaev:2010kp}
A.~Belyaev, M.~T. Frandsen, S.~Sarkar, and F.~Sannino, {\it {Mixed dark matter
  from technicolor}},  {\em Phys.Rev.} {\bf D83} (2011) 015007,
  [\href{http://xxx.lanl.gov/abs/1007.4839}{{\tt arXiv:1007.4839}}].

\bibitem{Lewis:2011zb}
R.~Lewis, C.~Pica, and F.~Sannino, {\it {Light Asymmetric Dark Matter on the
  Lattice: SU(2) Technicolor with Two Fundamental Flavors}},  {\em Phys.Rev.}
  {\bf D85} (2012) 014504, [\href{http://xxx.lanl.gov/abs/1109.3513}{{\tt
  arXiv:1109.3513}}].

\bibitem{Buckley:2012ky}
M.~R. Buckley and E.~T. Neil, {\it {Thermal Dark Matter from a Confining
  Sector}},  {\em Phys.Rev.} {\bf D87} (2013), no.~4 043510,
  [\href{http://xxx.lanl.gov/abs/1209.6054}{{\tt arXiv:1209.6054}}].

\bibitem{Hietanen:2012qd}
A.~Hietanen, C.~Pica, F.~Sannino, and U.~I. Sondergaard, {\it {Isotriplet Dark
  Matter on the Lattice: SO(4)-gauge theory with two Vector Wilson fermions}},
  {\em PoS} {\bf LATTICE2012} (2012) 065,
  [\href{http://xxx.lanl.gov/abs/1211.0142}{{\tt arXiv:1211.0142}}].

\bibitem{Hietanen:2012sz}
A.~Hietanen, C.~Pica, F.~Sannino, and U.~I. Sondergaard, {\it {Orthogonal
  Technicolor with Isotriplet Dark Matter on the Lattice}},  {\em Phys.Rev.}
  {\bf D87} (2013), no.~3 034508,
  [\href{http://xxx.lanl.gov/abs/1211.5021}{{\tt arXiv:1211.5021}}].

\bibitem{Appelquist:2013ms}
{\bf Lattice Strong Dynamics (LSD) Collaboration} Collaboration, T.~Appelquist
  {\em et.~al.}, {\it {Lattice calculation of composite dark matter form
  factors}},  {\em Phys.Rev.} {\bf D88} (2013), no.~1 014502,
  [\href{http://xxx.lanl.gov/abs/1301.1693}{{\tt arXiv:1301.1693}}].

\bibitem{Hietanen:2013fya}
A.~Hietanen, R.~Lewis, C.~Pica, and F.~Sannino, {\it {Composite Goldstone Dark
  Matter: Experimental Predictions from the Lattice}},
  \href{http://xxx.lanl.gov/abs/1308.4130}{{\tt arXiv:1308.4130}}.

\bibitem{Cline:2013zca}
J.~M. Cline, Z.~Liu, G.~Moore, and W.~Xue, {\it {Composite strongly interacting
  dark matter}},  \href{http://xxx.lanl.gov/abs/1312.3325}{{\tt
  arXiv:1312.3325}}.

\bibitem{Appelquist:2014dja}
T.~Appelquist, E.~Berkowitz, R.~C. Brower, M.~I. Buchoff, G.~T. Fleming, {\em
  et.~al.}, {\it {Composite bosonic baryon dark matter on the lattice: SU(4)
  baryon spectrum and the effective Higgs interaction}},
  \href{http://xxx.lanl.gov/abs/1402.6656}{{\tt arXiv:1402.6656}}.

\bibitem{Braaten:2013tza}
E.~Braaten and H.~W. Hammer, {\it {Universal Two-body Physics in Dark Matter
  near an S-wave Resonance}},  {\em Phys.Rev.} {\bf D88} (2013) 063511,
  [\href{http://xxx.lanl.gov/abs/1303.4682}{{\tt arXiv:1303.4682}}].

\bibitem{Laha:2013gva}
R.~Laha and E.~Braaten, {\it {Direct detection of dark matter in universal
  bound states}},  {\em Phys.Rev.} {\bf D89} (2014) 103510,
  [\href{http://xxx.lanl.gov/abs/1311.6386}{{\tt arXiv:1311.6386}}].

\bibitem{Beane:2012vq}
S.~Beane, E.~Chang, S.~Cohen, W.~Detmold, H.~Lin, {\em et.~al.}, {\it {Light
  Nuclei and Hypernuclei from Quantum Chromodynamics in the Limit of SU(3)
  Flavor Symmetry}},  {\em Phys.Rev.} {\bf D87} (2013), no.~3 034506,
  [\href{http://xxx.lanl.gov/abs/1206.5219}{{\tt arXiv:1206.5219}}].

\bibitem{Detmold:2012eu}
W.~Detmold and K.~Orginos, {\it {Nuclear correlation functions in lattice
  QCD}},  {\em Phys.Rev.} {\bf D87} (2013), no.~11 114512,
  [\href{http://xxx.lanl.gov/abs/1207.1452}{{\tt arXiv:1207.1452}}].

\bibitem{Yamazaki:2012hi}
T.~Yamazaki, K.-i. Ishikawa, Y.~Kuramashi, and A.~Ukawa, {\it {Helium nuclei,
  deuteron and dineutron in 2+1 flavor lattice QCD}},  {\em Phys.Rev.} {\bf
  D86} (2012) 074514, [\href{http://xxx.lanl.gov/abs/1207.4277}{{\tt
  arXiv:1207.4277}}].

\bibitem{Krnjaic:2014xza}
G.~Krnjaic and K.~Sigurdson, {\it {Big Bang Darkleosynthesis}},
  \href{http://xxx.lanl.gov/abs/1406.1171}{{\tt arXiv:1406.1171}}.

\bibitem{D'Eramo:2010ep}
F.~D'Eramo and J.~Thaler, {\it {Semi-annihilation of Dark Matter}},  {\em JHEP}
  {\bf 1006} (2010) 109, [\href{http://xxx.lanl.gov/abs/1003.5912}{{\tt
  arXiv:1003.5912}}].

\bibitem{D'Eramo:2011ec}
F.~D'Eramo, L.~Fei, and J.~Thaler, {\it {Dark Matter Assimilation into the
  Baryon Asymmetry}},  {\em JCAP} {\bf 1203} (2012) 010,
  [\href{http://xxx.lanl.gov/abs/1111.5615}{{\tt arXiv:1111.5615}}].

\bibitem{D'Eramo:2012rr}
F.~D'Eramo, M.~McCullough, and J.~Thaler, {\it {Multiple Gamma Lines from
  Semi-Annihilation}},  {\em JCAP} {\bf 1304} (2013) 030,
  [\href{http://xxx.lanl.gov/abs/1210.7817}{{\tt arXiv:1210.7817}}].

\bibitem{Belanger:2012vp}
G.~Belanger, K.~Kannike, A.~Pukhov, and M.~Raidal, {\it {Impact of
  semi-annihilations on dark matter phenomenology - an example of ZN symmetric
  scalar dark matter}},  {\em JCAP} {\bf 1204} (2012) 010,
  [\href{http://xxx.lanl.gov/abs/1202.2962}{{\tt arXiv:1202.2962}}].

\bibitem{Arina:2009uq}
C.~Arina, T.~Hambye, A.~Ibarra, and C.~Weniger, {\it {Intense Gamma-Ray Lines
  from Hidden Vector Dark Matter Decay}},  {\em JCAP} {\bf 1003} (2010) 024,
  [\href{http://xxx.lanl.gov/abs/0912.4496}{{\tt arXiv:0912.4496}}].

\bibitem{Hambye:2009fg}
T.~Hambye and M.~H. Tytgat, {\it {Confined hidden vector dark matter}},  {\em
  Phys.Lett.} {\bf B683} (2010) 39--41,
  [\href{http://xxx.lanl.gov/abs/0907.1007}{{\tt arXiv:0907.1007}}].

\bibitem{Hambye:2008bq}
T.~Hambye, {\it {Hidden vector dark matter}},  {\em JHEP} {\bf 0901} (2009)
  028, [\href{http://xxx.lanl.gov/abs/0811.0172}{{\tt arXiv:0811.0172}}].

\bibitem{Bertolini:2012gu}
D.~Bertolini and M.~McCullough, {\it {The Social Higgs}},  {\em JHEP} {\bf
  1212} (2012) 118, [\href{http://xxx.lanl.gov/abs/1207.4209}{{\tt
  arXiv:1207.4209}}].

\bibitem{Belanger:2013kya}
G.~Belanger, B.~Dumont, U.~Ellwanger, J.~Gunion, and S.~Kraml, {\it {Status of
  invisible Higgs decays}},  {\em Phys.Lett.} {\bf B723} (2013) 340--347,
  [\href{http://xxx.lanl.gov/abs/1302.5694}{{\tt arXiv:1302.5694}}].

\bibitem{Giardino:2013bma}
P.~P. Giardino, K.~Kannike, I.~Masina, M.~Raidal, and A.~Strumia, {\it {The
  universal Higgs fit}},  {\em JHEP} {\bf 1405} (2014) 046,
  [\href{http://xxx.lanl.gov/abs/1303.3570}{{\tt arXiv:1303.3570}}].

\bibitem{Ellis:2013lra}
J.~Ellis and T.~You, {\it {Updated Global Analysis of Higgs Couplings}},  {\em
  JHEP} {\bf 1306} (2013) 103, [\href{http://xxx.lanl.gov/abs/1303.3879}{{\tt
  arXiv:1303.3879}}].

\bibitem{Peskin:1980gc}
M.~E. Peskin, {\it {The Alignment of the Vacuum in Theories of Technicolor}},
  {\em Nucl.Phys.} {\bf B175} (1980) 197--233.

\bibitem{Preskill:1980mz}
J.~Preskill, {\it {Subgroup Alignment in Hypercolor Theories}},  {\em
  Nucl.Phys.} {\bf B177} (1981) 21--59.

\bibitem{Kosower:1984aw}
D.~Kosower, {\it {SYMMETRY BREAKING PATTERNS IN PSEUDOREAL AND REAL GAUGE
  THEORIES}},  {\em Phys.Lett.} {\bf B144} (1984) 215--216.

\bibitem{Coleman:1969sm}
S.~R. Coleman, J.~Wess, and B.~Zumino, {\it {Structure of phenomenological
  Lagrangians. 1.}},  {\em Phys.Rev.} {\bf 177} (1969) 2239--2247.

\bibitem{Callan:1969sn}
J.~Callan, Curtis~G., S.~R. Coleman, J.~Wess, and B.~Zumino, {\it {Structure of
  phenomenological Lagrangians. 2.}},  {\em Phys.Rev.} {\bf 177} (1969)
  2247--2250.

\bibitem{Jenkins:1995vb}
E.~E. Jenkins, A.~V. Manohar, and M.~B. Wise, {\it {Chiral perturbation theory
  for vector mesons}},  {\em Phys.Rev.Lett.} {\bf 75} (1995) 2272--2275,
  [\href{http://xxx.lanl.gov/abs/hep-ph/9506356}{{\tt hep-ph/9506356}}].

\bibitem{Griest:1990kh}
K.~Griest and D.~Seckel, {\it {Three exceptions in the calculation of relic
  abundances}},  {\em Phys.Rev.} {\bf D43} (1991) 3191--3203.

\bibitem{Kolb:1990vq}
E.~W. Kolb and M.~S. Turner, {\it {The Early Universe}},  {\em Front.Phys.}
  {\bf 69} (1990) 1--547.

\bibitem{Hall:2009bx}
L.~J. Hall, K.~Jedamzik, J.~March-Russell, and S.~M. West, {\it {Freeze-In
  Production of FIMP Dark Matter}},  {\em JHEP} {\bf 1003} (2010) 080,
  [\href{http://xxx.lanl.gov/abs/0911.1120}{{\tt arXiv:0911.1120}}].

\bibitem{Gelmini:1986zz}
G.~Gelmini, L.~J. Hall, and M.~Lin, {\it {What Is the Cosmion?}},  {\em
  Nucl.Phys.} {\bf B281} (1987) 726.

\bibitem{Chivukula:1989qb}
R.~S. Chivukula and T.~P. Walker, {\it {TECHNICOLOR COSMOLOGY}},  {\em
  Nucl.Phys.} {\bf B329} (1990) 445.

\bibitem{Kaplan:1991ah}
D.~B. Kaplan, {\it {A Single explanation for both the baryon and dark matter
  densities}},  {\em Phys.Rev.Lett.} {\bf 68} (1992) 741--743.

\bibitem{Thomas:1995ze}
S.~D. Thomas, {\it {Baryons and dark matter from the late decay of a
  supersymmetric condensate}},  {\em Phys.Lett.} {\bf B356} (1995) 256--263,
  [\href{http://xxx.lanl.gov/abs/hep-ph/9506274}{{\tt hep-ph/9506274}}].

\bibitem{Hooper:2004dc}
D.~Hooper, J.~March-Russell, and S.~M. West, {\it {Asymmetric sneutrino dark
  matter and the Omega(b) / Omega(DM) puzzle}},  {\em Phys.Lett.} {\bf B605}
  (2005) 228--236, [\href{http://xxx.lanl.gov/abs/hep-ph/0410114}{{\tt
  hep-ph/0410114}}].

\bibitem{Kitano:2004sv}
R.~Kitano and I.~Low, {\it {Dark matter from baryon asymmetry}},  {\em
  Phys.Rev.} {\bf D71} (2005) 023510,
  [\href{http://xxx.lanl.gov/abs/hep-ph/0411133}{{\tt hep-ph/0411133}}].

\bibitem{Agashe:2004bm}
K.~Agashe and G.~Servant, {\it {Baryon number in warped GUTs: Model building
  and (dark matter related) phenomenology}},  {\em JCAP} {\bf 0502} (2005) 002,
  [\href{http://xxx.lanl.gov/abs/hep-ph/0411254}{{\tt hep-ph/0411254}}].

\bibitem{Cosme:2005sb}
N.~Cosme, L.~Lopez~Honorez, and M.~H. Tytgat, {\it {Leptogenesis and dark
  matter related?}},  {\em Phys.Rev.} {\bf D72} (2005) 043505,
  [\href{http://xxx.lanl.gov/abs/hep-ph/0506320}{{\tt hep-ph/0506320}}].

\bibitem{Farrar:2005zd}
G.~R. Farrar and G.~Zaharijas, {\it {Dark matter and the baryon asymmetry}},
  {\em Phys.Rev.Lett.} {\bf 96} (2006) 041302,
  [\href{http://xxx.lanl.gov/abs/hep-ph/0510079}{{\tt hep-ph/0510079}}].

\bibitem{Suematsu:2005kp}
D.~Suematsu, {\it {Nonthermal production of baryon and dark matter}},  {\em
  Astropart.Phys.} {\bf 24} (2006) 511--519,
  [\href{http://xxx.lanl.gov/abs/hep-ph/0510251}{{\tt hep-ph/0510251}}].

\bibitem{Tytgat:2006wy}
M.~H. Tytgat, {\it {Relating leptogenesis and dark matter}},
  \href{http://xxx.lanl.gov/abs/hep-ph/0606140}{{\tt hep-ph/0606140}}.

\bibitem{Banks:2006xr}
T.~Banks, S.~Echols, and J.~Jones, {\it {Baryogenesis, dark matter and the
  Pentagon}},  {\em JHEP} {\bf 0611} (2006) 046,
  [\href{http://xxx.lanl.gov/abs/hep-ph/0608104}{{\tt hep-ph/0608104}}].

\bibitem{Kitano:2008tk}
R.~Kitano, H.~Murayama, and M.~Ratz, {\it {Unified origin of baryons and dark
  matter}},  {\em Phys.Lett.} {\bf B669} (2008) 145--149,
  [\href{http://xxx.lanl.gov/abs/0807.4313}{{\tt arXiv:0807.4313}}].

\bibitem{Kaplan:2009ag}
D.~E. Kaplan, M.~A. Luty, and K.~M. Zurek, {\it {Asymmetric Dark Matter}},
  {\em Phys.Rev.} {\bf D79} (2009) 115016,
  [\href{http://xxx.lanl.gov/abs/0901.4117}{{\tt arXiv:0901.4117}}].

\bibitem{Petraki:2013wwa}
K.~Petraki and R.~R. Volkas, {\it {Review of asymmetric dark matter}},  {\em
  Int.J.Mod.Phys.} {\bf A28} (2013) 1330028,
  [\href{http://xxx.lanl.gov/abs/1305.4939}{{\tt arXiv:1305.4939}}].

\bibitem{Zurek:2013wia}
K.~M. Zurek, {\it {Asymmetric Dark Matter: Theories, Signatures, and
  Constraints}},  {\em Phys.Rept.} {\bf 537} (2014) 91--121,
  [\href{http://xxx.lanl.gov/abs/1308.0338}{{\tt arXiv:1308.0338}}].

\bibitem{Buckley:2010ui}
M.~R. Buckley and L.~Randall, {\it {Xogenesis}},  {\em JHEP} {\bf 1109} (2011)
  009, [\href{http://xxx.lanl.gov/abs/1009.0270}{{\tt arXiv:1009.0270}}].

\bibitem{MarchRussell:2011fi}
J.~March-Russell and M.~McCullough, {\it {Asymmetric Dark Matter via
  Spontaneous Co-Genesis}},  {\em JCAP} {\bf 1203} (2012) 019,
  [\href{http://xxx.lanl.gov/abs/1106.4319}{{\tt arXiv:1106.4319}}].

\bibitem{Graesser:2011wi}
M.~L. Graesser, I.~M. Shoemaker, and L.~Vecchi, {\it {Asymmetric WIMP dark
  matter}},  {\em JHEP} {\bf 1110} (2011) 110,
  [\href{http://xxx.lanl.gov/abs/1103.2771}{{\tt arXiv:1103.2771}}].

\bibitem{Daylan:2014rsa}
T.~Daylan, D.~P. Finkbeiner, D.~Hooper, T.~Linden, S.~K.~N. Portillo, {\em
  et.~al.}, {\it {The Characterization of the Gamma-Ray Signal from the Central
  Milky Way: A Compelling Case for Annihilating Dark Matter}},
  \href{http://xxx.lanl.gov/abs/1402.6703}{{\tt arXiv:1402.6703}}.

\bibitem{Goodenough:2009gk}
L.~Goodenough and D.~Hooper, {\it {Possible Evidence For Dark Matter
  Annihilation In The Inner Milky Way From The Fermi Gamma Ray Space
  Telescope}},  \href{http://xxx.lanl.gov/abs/0910.2998}{{\tt
  arXiv:0910.2998}}.

\bibitem{Hooper:2010mq}
D.~Hooper and L.~Goodenough, {\it {Dark Matter Annihilation in The Galactic
  Center As Seen by the Fermi Gamma Ray Space Telescope}},  {\em Phys.Lett.}
  {\bf B697} (2011) 412--428, [\href{http://xxx.lanl.gov/abs/1010.2752}{{\tt
  arXiv:1010.2752}}].

\bibitem{Hooper:2011ti}
D.~Hooper and T.~Linden, {\it {On The Origin Of The Gamma Rays From The
  Galactic Center}},  {\em Phys.Rev.} {\bf D84} (2011) 123005,
  [\href{http://xxx.lanl.gov/abs/1110.0006}{{\tt arXiv:1110.0006}}].

\bibitem{Abazajian:2012pn}
K.~N. Abazajian and M.~Kaplinghat, {\it {Detection of a Gamma-Ray Source in the
  Galactic Center Consistent with Extended Emission from Dark Matter
  Annihilation and Concentrated Astrophysical Emission}},  {\em Phys.Rev.} {\bf
  D86} (2012) 083511, [\href{http://xxx.lanl.gov/abs/1207.6047}{{\tt
  arXiv:1207.6047}}].

\bibitem{Hooper:2013rwa}
D.~Hooper and T.~R. Slatyer, {\it {Two Emission Mechanisms in the Fermi
  Bubbles: A Possible Signal of Annihilating Dark Matter}},  {\em Phys.Dark
  Univ.} {\bf 2} (2013) 118--138,
  [\href{http://xxx.lanl.gov/abs/1302.6589}{{\tt arXiv:1302.6589}}].

\bibitem{Gordon:2013vta}
C.~Gordon and O.~Macias, {\it {Dark Matter and Pulsar Model Constraints from
  Galactic Center Fermi-LAT Gamma Ray Observations}},  {\em Phys.Rev.} {\bf
  D88} (2013) 083521, [\href{http://xxx.lanl.gov/abs/1306.5725}{{\tt
  arXiv:1306.5725}}].

\bibitem{Huang:2013pda}
W.-C. Huang, A.~Urbano, and W.~Xue, {\it {Fermi Bubbles under Dark Matter
  Scrutiny. Part I: Astrophysical Analysis}},
  \href{http://xxx.lanl.gov/abs/1307.6862}{{\tt arXiv:1307.6862}}.

\bibitem{Abazajian:2014fta}
K.~N. Abazajian, N.~Canac, S.~Horiuchi, and M.~Kaplinghat, {\it {Astrophysical
  and Dark Matter Interpretations of Extended Gamma Ray Emission from the
  Galactic Center}},  \href{http://xxx.lanl.gov/abs/1402.4090}{{\tt
  arXiv:1402.4090}}.

\bibitem{Carlson:2014cwa}
E.~Carlson and S.~Profumo, {\it {Cosmic Ray Protons in the Inner Galaxy and the
  Galactic Center Gamma-Ray Excess}},
  \href{http://xxx.lanl.gov/abs/1405.7685}{{\tt arXiv:1405.7685}}.

\bibitem{Petrovic:2014uda}
J.~Petrovic, P.~D. Serpico, and G.~Zaharijas, {\it {Galactic Center gamma-ray
  "excess" from an active past of the Galactic Centre?}},
  \href{http://xxx.lanl.gov/abs/1405.7928}{{\tt arXiv:1405.7928}}.

\bibitem{Cirelli:2010xx}
M.~Cirelli, G.~Corcella, A.~Hektor, G.~Hutsi, M.~Kadastik, {\em et.~al.}, {\it
  {PPPC 4 DM ID: A Poor Particle Physicist Cookbook for Dark Matter Indirect
  Detection}},  {\em JCAP} {\bf 1103} (2011) 051,
  [\href{http://xxx.lanl.gov/abs/1012.4515}{{\tt arXiv:1012.4515}}].

\bibitem{Cirelli:2013mqa}
M.~Cirelli, P.~D. Serpico, and G.~Zaharijas, {\it {Bremsstrahlung gamma rays
  from light Dark Matter}},  {\em JCAP} {\bf 1311} (2013) 035,
  [\href{http://xxx.lanl.gov/abs/1307.7152}{{\tt arXiv:1307.7152}}].

\bibitem{Navarro:1995iw}
J.~F. Navarro, C.~S. Frenk, and S.~D. White, {\it {The Structure of cold dark
  matter halos}},  {\em Astrophys.J.} {\bf 462} (1996) 563--575,
  [\href{http://xxx.lanl.gov/abs/astro-ph/9508025}{{\tt astro-ph/9508025}}].

\bibitem{Boehm:2014bia}
C.~Boehm, M.~J. Dolan, and C.~McCabe, {\it {A weighty interpretation of the
  Galactic Centre excess}},  \href{http://xxx.lanl.gov/abs/1404.4977}{{\tt
  arXiv:1404.4977}}.

\bibitem{Buckley:2011ye}
M.~R. Buckley and S.~Profumo, {\it {Regenerating a Symmetry in Asymmetric Dark
  Matter}},  {\em Phys.Rev.Lett.} {\bf 108} (2012) 011301,
  [\href{http://xxx.lanl.gov/abs/1109.2164}{{\tt arXiv:1109.2164}}].

\bibitem{Cirelli:2011ac}
M.~Cirelli, P.~Panci, G.~Servant, and G.~Zaharijas, {\it {Consequences of
  DM/antiDM Oscillations for Asymmetric WIMP Dark Matter}},  {\em JCAP} {\bf
  1203} (2012) 015, [\href{http://xxx.lanl.gov/abs/1110.3809}{{\tt
  arXiv:1110.3809}}].

\bibitem{Tulin:2012re}
S.~Tulin, H.-B. Yu, and K.~M. Zurek, {\it {Oscillating Asymmetric Dark
  Matter}},  {\em JCAP} {\bf 1205} (2012) 013,
  [\href{http://xxx.lanl.gov/abs/1202.0283}{{\tt arXiv:1202.0283}}].

\bibitem{Okada:2012rm}
N.~Okada and O.~Seto, {\it {Originally Asymmetric Dark Matter}},  {\em
  Phys.Rev.} {\bf D86} (2012) 063525,
  [\href{http://xxx.lanl.gov/abs/1205.2844}{{\tt arXiv:1205.2844}}].

\bibitem{Hardy:2014dea}
E.~Hardy, R.~Lasenby, and J.~Unwin, {\it {Annihilation Signals from Asymmetric
  Dark Matter}},  \href{http://xxx.lanl.gov/abs/1402.4500}{{\tt
  arXiv:1402.4500}}.

\bibitem{Spergel:1984re}
D.~Spergel and W.~Press, {\it {Effect of hypothetical, weakly interacting,
  massive particles on energy transport in the solar interior}},  {\em
  Astrophys.J.} {\bf 294} (1985) 663--673.

\bibitem{Press:1985ug}
W.~H. Press and D.~N. Spergel, {\it {Capture by the sun of a galactic
  population of weakly interacting massive particles}},  {\em Astrophys.J.}
  {\bf 296} (1985) 679--684.

\bibitem{Silk:1985ax}
J.~Silk, K.~A. Olive, and M.~Srednicki, {\it {The Photino, the Sun and
  High-Energy Neutrinos}},  {\em Phys.Rev.Lett.} {\bf 55} (1985) 257--259.

\bibitem{Srednicki:1986vj}
M.~Srednicki, K.~A. Olive, and J.~Silk, {\it {High-Energy Neutrinos from the
  Sun and Cold Dark Matter}},  {\em Nucl.Phys.} {\bf B279} (1987) 804.

\bibitem{Gould:1987ju}
A.~Gould, {\it {{WIMP} Distribution in and Evaporation From the Sun}},  {\em
  Astrophys.J.} {\bf 321} (1987) 560.

\bibitem{Gould:1987ir}
A.~Gould, {\it {Resonant Enhancements in WIMP Capture by the Earth}},  {\em
  Astrophys.J.} {\bf 321} (1987) 571.

\bibitem{Gould:1987ww}
A.~Gould, {\it {Direct and Indirect Capture of Wimps by the Earth}},  {\em
  Astrophys.J.} {\bf 328} (1988) 919--939.

\bibitem{Steigman:1997vs}
G.~Steigman, C.~Sarazin, H.~Quintana, and J.~Faulkner, {\it {Dynamical
  interactions and astrophysical effects of stable heavy neutrinos}}, .

\bibitem{Bottino:2002pd}
A.~Bottino, G.~Fiorentini, N.~Fornengo, B.~Ricci, S.~Scopel, {\em et.~al.},
  {\it {Does solar physics provide constraints to weakly interacting massive
  particles?}},  {\em Phys.Rev.} {\bf D66} (2002) 053005,
  [\href{http://xxx.lanl.gov/abs/hep-ph/0206211}{{\tt hep-ph/0206211}}].

\bibitem{Moskalenko:2007ak}
I.~V. Moskalenko and L.~L. Wai, {\it {Dark matter burners}},  {\em
  Astrophys.J.} {\bf 659} (2007) L29--L32,
  [\href{http://xxx.lanl.gov/abs/astro-ph/0702654}{{\tt astro-ph/0702654}}].

\bibitem{Bertone:2007ae}
G.~Bertone and M.~Fairbairn, {\it {Compact Stars as Dark Matter Probes}},  {\em
  Phys.Rev.} {\bf D77} (2008) 043515,
  [\href{http://xxx.lanl.gov/abs/0709.1485}{{\tt arXiv:0709.1485}}].

\bibitem{McCullough:2010ai}
M.~McCullough and M.~Fairbairn, {\it {Capture of Inelastic Dark Matter in White
  Dwarves}},  {\em Phys.Rev.} {\bf D81} (2010) 083520,
  [\href{http://xxx.lanl.gov/abs/1001.2737}{{\tt arXiv:1001.2737}}].

\bibitem{Frandsen:2010yj}
M.~T. Frandsen and S.~Sarkar, {\it {Asymmetric dark matter and the Sun}},  {\em
  Phys.Rev.Lett.} {\bf 105} (2010) 011301,
  [\href{http://xxx.lanl.gov/abs/1003.4505}{{\tt arXiv:1003.4505}}].

\bibitem{Kouvaris:2010jy}
C.~Kouvaris and P.~Tinyakov, {\it {Constraining Asymmetric Dark Matter through
  observations of compact stars}},  {\em Phys.Rev.} {\bf D83} (2011) 083512,
  [\href{http://xxx.lanl.gov/abs/1012.2039}{{\tt arXiv:1012.2039}}].

\bibitem{McDermott:2011jp}
S.~D. McDermott, H.-B. Yu, and K.~M. Zurek, {\it {Constraints on Scalar
  Asymmetric Dark Matter from Black Hole Formation in Neutron Stars}},  {\em
  Phys.Rev.} {\bf D85} (2012) 023519,
  [\href{http://xxx.lanl.gov/abs/1103.5472}{{\tt arXiv:1103.5472}}].

\bibitem{Kouvaris:2011fi}
C.~Kouvaris and P.~Tinyakov, {\it {Excluding Light Asymmetric Bosonic Dark
  Matter}},  {\em Phys.Rev.Lett.} {\bf 107} (2011) 091301,
  [\href{http://xxx.lanl.gov/abs/1104.0382}{{\tt arXiv:1104.0382}}].

\bibitem{Iocco:2012wk}
F.~Iocco, M.~Taoso, F.~Leclercq, and G.~Meynet, {\it {Main sequence stars with
  asymmetric dark matter}},  {\em Phys.Rev.Lett.} {\bf 108} (2012) 061301,
  [\href{http://xxx.lanl.gov/abs/1201.5387}{{\tt arXiv:1201.5387}}].

\bibitem{Lopes:2012af}
I.~Lopes and J.~Silk, {\it {Solar constraints on asymmetric dark matter}},
  {\em Astrophys.J.} {\bf 757} (2012) 130,
  [\href{http://xxx.lanl.gov/abs/1209.3631}{{\tt arXiv:1209.3631}}].

\bibitem{Bell:2013xk}
N.~F. Bell, A.~Melatos, and K.~Petraki, {\it {Realistic neutron star
  constraints on bosonic asymmetric dark matter}},  {\em Phys.Rev.} {\bf D87}
  (2013), no.~12 123507, [\href{http://xxx.lanl.gov/abs/1301.6811}{{\tt
  arXiv:1301.6811}}].

\bibitem{Huang:2013xfa}
J.~Huang and Y.~Zhao, {\it {Dark Matter Induced Nucleon Decay: Model and
  Signatures}},  {\em JHEP} {\bf 1402} (2014) 077,
  [\href{http://xxx.lanl.gov/abs/1312.0011}{{\tt arXiv:1312.0011}}].

\bibitem{Agashe:2014yua}
K.~Agashe, Y.~Cui, L.~Necib, and J.~Thaler, {\it {(In)direct Detection of
  Boosted Dark Matter}},  \href{http://xxx.lanl.gov/abs/1405.7370}{{\tt
  arXiv:1405.7370}}.

\end{thebibliography}\endgroup

\end{document}